\documentclass[a4paper,11pt]{article}
\pdfoutput=1 

\usepackage{jcappub} 

\usepackage[T1]{fontenc} 

\usepackage{graphicx}
\usepackage{caption}
\usepackage{subcaption}
\usepackage{multirow}
\usepackage{diagbox}
\usepackage{adjustbox}
\usepackage{hyperref}
\usepackage[utf8]{inputenc}
\usepackage{ulem}

\title{\boldmath Ultralight DM bosons with an axion-like potential:  scale-dependent constraints revisited. }

\author[a,b,1]{Francisco X. Linares Cede\~no,\note{Corresponding author.}}
\author[c,d]{Alma X. Gonz\'alez-Morales}
\author[c]{and L. Arturo Ure\~na-L\'opez,}

\affiliation[a]{Instituto de F\'isica y Matem\'aticas, Universidad Michoacana de San Nicol\'as de Hidalgo, Edificio C-3, Ciudad Universitaria, CP. 58040 Morelia, Michoac\'an, M\'exico.}
\affiliation[b]{Mesoamerican Centre for Theoretical Physics, Universidad Autónoma de Chiapas, Carretera Zapata Km 4, Real del Bosque (Terán), 29040, Tuxla Gutierrez, Chiapas, México.}
\affiliation[c]{Departamento de F\'isica, DCI, Campus Le\'on, Universidad de
Guanajuato, 37150, Le\'on, Guanajuato, M\'exico.}
\affiliation[d]{Consejo Nacional de Ciencia y Tecnolog\'ia,
Av. Insurgentes Sur 1582. Colonia Cr\'edito Constructor, Del. Benito   Ju\'arez C.P. 03940, M\'exico D.F. M\'exico}

\emailAdd{francisco.linares@umich.mx}
\emailAdd{lurena@ugto.mx}
\emailAdd{alma.gonzalez@fisica.ugto.mx}

\abstract{A scalar field $\phi$ endowed with a trigonometric potential has been proposed to play the role of Dark Matter. A deep study of the cosmological evolution of linear perturbations, and its comparison to the  \textit{Cold Dark Matter} (CDM) and \textit{Fuzzy Dark Matter} (FDM) cases (scalar field with quadratic potential), reveals an enhancement in the amplitude of the mass power spectrum for large wave numbers due to the non--linearity of the axion--like potential. For the first time, we study the scale--dependence on physical quantities such as the growth factor $D_k$, the velocity growth factor $f_k$, and $f_k \sigma_8$. We found that for $z<10$, all these quantities recover the CDM evolution, whereas for high redshift there is a clear distinction between each model (FDM case, and axion--like potential) depending on the wavenumber $k$ and on the decay parameter of the axion-like potential as well. A semi--analytical Halo Mass Function is also revisited, finding a suppression of the number of low mass halos, as in the FDM case, but with a small increment in the amplitude of the variance and halo mass function due to the non--linearity of the axion--like potential. Finally, we present constraints on the axion mass and the axion decay parameter by using data of the Planck Collaboration 2018 and Lyman-$\alpha$ forest.}

\begin{document}
\maketitle
\flushbottom

\section{Introduction}
\label{sec:intro}

One of the open problems of modern physics concerns the existence of Dark Matter (DM). At present we have several observations indicating that such a component of matter exists~\cite{Roos:2012cc,Buckley:2017ijx,Salucci:2018eie,Bertone:2019irm,Rubakov:2019lyf}, and that it is most likely the main agent driving the formation of structure. The most successful model describing this unknown component of matter is called \textit{Cold Dark Matter} (CDM), and consists of a pressureless fluid of particles that interacts mostly gravitationally with other components of matter~\cite{liddle1993cold,Eisenstein:1997jh}. Although the CDM model is so far in good agreement with most of the cosmological observations, DM nature is still a mystery.  It is well known that there are some differences at small scales between astrophysical observations and numerical simulations based on CDM \cite{Klypin:1999uc,bullock20133,patrick2002globular,Penarrubia:2012bb,
de2009core,marsh2015axion,li2009cusp,Jiang:2015vra,Garrison-Kimmel:2014vqa,Weinberg:2013aya,Pawlowski:2015qta,DelPopolo:2016emo,Bullock:2017xww,Kim:2017iwr,Brooks:2018ktu,Fitts:2018ycl}. These differences may be due to the lack of information about astrophysical processes of galactic substructures and baryonic physics, but they could also be a manifestation of the still unknown properties of the DM field. Hence, if it is the case that DM is the main responsible for the process of structure and substructure formation, then it is important to explore and analyze other DM candidates that could offer a better description of the structures at such small scales. With many different models in the literature, it is important that any given model under study predicts observables accurately so that comparison against observations are meaningful.

In this context, models of Scalar Field Dark Matter (SFDM) have gained great relevance in modern cosmology by becoming a promising candidate to describe the DM as well, maybe even better, than CDM. While the implementation of scalar fields in cosmology has historically its origins in inflationary models of the early Universe\cite{Guth:1980zm,Linde:1981mu}, and as candidates to explain the accelerated expansion of the Universe at late times \cite{Caldwell:1998je,Ratra:1987rm,Peebles:1987ek,Wetterich:1987fm,Boyle:2001du}, scalar fields also possess interesting properties that make them suitable as DM models.

A compelling DM candidate that has been vastly investigated, and which is the one of interest in this work, is the \textit{axion}, a scalar field originally proposed to solve the Strong CP problem in QCD \cite{PhysRevLett.38.1440,PhysRevD.16.1791,PhysRevLett.40.223,PhysRevLett.40.279}, and which origin can be given within a more fundamental theory such as String Theory \cite{witten1984some,svrcek2006axions,cicoli2012type,cicoli2014global,ringwald2014searching,ahn2016qcd,Visinelli:2018utg}. Several models involving axions and \textit{axion-like particles} have surged as possible source for DM \cite{Preskill:1982cy,Dine:1982ah,abbott1983cosmological,sikivie1994dark,masso2003axions,raffelt2007axions,Sikivie:2006ni,hwang2009axion,ringwald2014axions,Gonzales-Morales:2016mkl,DAmico:2016jbm,Hui2016:1610.08297v2,Ballesteros:2016euj,Hlozek2017:1708.05681v1,Poddar:2019zoe}, and several experiments such as ADMX \cite{asztalos2004improved}, SOLAX\cite{avignone1998experimental}, DAMA \cite{bernabei2001search}, COSME \cite{morales2002particle}, CAST \cite{zioutas2005first} and the ORGAN experiment~\cite{McAllister:2017lkb}, are trying to hunt directly this elusive kind of particles (other possible ways of detection can be seen in \cite{PhysRevLett.104.041301,khmelnitsky2014pulsar,Aoki:2016kwl,sikivie1983experimental,bradley2003microwave,Brito:2017zvb,Brito:2017wnc,Li:2016mmc,Day:2018ckv}). The lighter axion in QCD has masses of around $\mu $eV, while for axion--like particles the mass lies within the range of $10^{-26}$eV$<m_\phi<10^{-18}$eV, which is the reason why the latter are also known in the literature as \textit{ultralight axions}. An important feature of this DM candidate is that it can give rise to \textit{Bose-Einstein condensates} through a phase transition \cite{UrenaLopez:2008zh,sin1994late,ji1994late,hu2000fuzzy,harko2011cosmological,PhysRevLett.117.121801,lundgren2010lukewarm,Eby:2017azn}, and it can form \textit{caustics} as well \cite{Sikivie:2013joa,Sikivie1998139,duffy2008caustic,erken2012cosmic,sikivie2011emerging,Sikivie:2009qn,duffy2006high,Banik:2017ygz}.
Thus, axions and axion-like particles are very well motivated DM candidates from the theoretical point of view.

Axion models in which the scalar field potential includes only the quadratic term, usually referred as free case or fuzzy dark matter (FDM), have been extensively studied in the literature \cite{Marsh:2015xka,Marsh:2010wq,Marsh:2012nm,Marsh:2013ywa,Hlozek:2014lca,Hui:2016ltb,Diez-Tejedor:2017ivd,Ferreira:2020fam}. We will refer to it as the FDM case from now on. However, such models do not capture all the implications that arise when including a full axion potential.  

In this work we will focus on a model that incorporates a trigonometric potential that is typical in axion studies, defined by 

\begin{equation}
  V(\phi) = m^2_\phi f^2_\phi \left[ 1+\cos\left( \phi/f_\phi \right) \right] \, ,   \label{eq:0}
\end{equation}
Here, $m_\phi$ is the axion mass, $f_\phi$ is the axion decay constant, and the two together $m^2_\phi f^2_\phi$ make up the height of the potential. In typical axion models, there is a relationship between the mass and the decay constant in which they are inversely proportional to each other, in particular for axions coming from M-Theory and Type IIB string theory \cite{svrcek2006axions,arvanitaki2010string,samir2010m}, where the decay constant is of the order of $10^{17} \, \mathrm{GeV}$. For the purposes of this paper, both parameters will be considered as independent one of each other.

The choice of the potential in Eq.~\eqref{eq:0} codifies the shift symmetry of the axion field, and our main aim is to analyze in detail the cosmological implications arising from the non--linearity of such potential. Previous works have shown some semi-analytical treatment \cite{Diez-Tejedor:2017ivd,Zhang:2017flu}, while a first attempt to a full analysis was presented in~\cite{Cedeno:2017sou}. The effects on the Cosmic Microwave Background (CMB) radiation and Mass Power Spectrum (MPS) of such anharmonic potential, but considering different exponents $[1-\cos(\phi/f_\phi)]^n$ with $n = 1, 2, 3$, have been studied in~\cite{Poulin:2018dzj}. As we will show in the present work, when considering extreme values of the decay constant with the potential~\eqref{eq:0}, it is possible to quantify deviations from the FDM case, regarding the structure formation at linear regime, like the enhancement of the MPS at small scales reported in \cite{Zhang:2017flu,Cedeno:2017sou}, as well as to analyze implications for other observables.

An outline of this work is as follows. In Section~\ref{blp} we study the cosmological background evolution and linear perturbations regime, by means of establishing new variables and a dynamical system that lead us to a generalization of the fluid equations. Using an amended version of the Boltzmann code \texttt{CLASS}~\cite{Lesgourgues:2011re}, we track the evolution and growth of the perturbations, with which we give a detailed analysis of the tachyonic instability suffered by the density perturbations, and show that only a set of wavenumbers corresponding to small scales are affected by such instability. 

The matter and temperature power spectra that arise from the axion model are presented in Section~\ref{sec:observables}, and we use them to impose some bounds on the free parameters of the model: the axion mass $m_a$ and the decay parameter $f_a$ mentioned above. We also make a qualitative assessment of how the Lyman-$\alpha$ 1D mass power spectrum could constrain the parameters of our model. We observe that while the FDM model with masses $m_{\phi} \leq 10^{-22}$eV  is ruled out, it is possible for the axion field to pass the constraints if endowed with the trigonometric potential~\eqref{eq:0}. 

Motivated by the characteristic cut-off that this model presents in the MPS, in Section~\ref{sec:halos} we define both the growth factor $D_k$ and the velocity growth factor $f_k$, not only as a function of the scale factor but also with their dependence on the length scale. We then analyze the combination $\left[f_k\sigma_8\right](z)$ and the semi-analytical Halo Mass Function (HMF), which, like in the case of the MPS, it shows an enhancement in its amplitude when considering the potential~\eqref{eq:0}. Finally, in Section~\ref{sec:conclusions}, we give some conclusions and perspectives for future work.

\section{Background and Linear Perturbations Dynamics}\label{blp}
In this section we show the dynamical equations for the evolution of both, background and linear perturbations of the axion model~\eqref{eq:0}. Following previous work~\cite{Cedeno:2017sou,Urena-Lopez:2015gur}, we rewrite these equations as a dynamical system and then, by a polar change of variables, we obtain a set of first order differential equations which is more appropriate for numerical studies of ultra--light axions than using directly the field equations.

\subsection{Background Evolution}
The Einstein-Klein-Gordon equations for a minimally-coupled scalar field $\phi$ endowed with a generic potential $V(\phi)$, in a Friedmann-Robertson-Walker spacetime with null spatial curvature are given by
\begin{subequations}
\label{eq:2}
  \begin{eqnarray}
    H^2 &=& \frac{\kappa^2}{3} \left( \sum_j \rho_j +
      \rho_\phi \right) \, , 
    \dot{H} = - \frac{\kappa^2}{2} \left[ \sum_j (\rho_j +
      p_j ) + (\rho_\phi + p_\phi) \right] \, , \label{eq:2a} \\
    \dot{\rho}_j &=& - 3 H (\rho_j + p_j ) \,
    , \quad \quad
    \ddot{\phi} = -3 H \dot{\phi} - \frac{dV(\phi)}{d\phi}  \, , \label{eq:2b}
  \end{eqnarray}
\end{subequations}
where $\kappa^2 = 8\pi G$, a dot denotes derivative with respect to cosmic time $t$, and $H$ is the Hubble parameter. Also, the scalar field energy density $\rho_\phi$ and pressure $p_\phi$ are given by the canonical expressions:
\begin{equation}
  \rho_\phi = \frac{1}{2} \dot{\phi}^2 + V(\phi) \, ,
  \quad p_\phi = \frac{1}{2} \dot{\phi}^2 - V(\phi) \, .
\end{equation}

To transform the Klein-Gordon (KG) equation~\eqref{eq:2b}, we define a new set of polar variables based on previous works~\cite{Copeland:1997et,Urena-Lopez:2015odd,Roy:2018nce}, that for the particular case of potential~\eqref{eq:0} read\footnote{The range for the axion field is $\phi/f_\phi = [0,2\pi]$, with the minimum of the potential at $\phi/f_\phi = \pi$, and correspondingly the ranges of the new variables are $\theta = [0,\infty)$, $\Omega_\phi = [0,1]$, and $y_1 = [0,\infty)$. Notice that the actual field dynamics is hidden in the new variables, but it can be easily recovered from Eqs.~\eqref{eq:defs}. For instance, one can show that $\cot (\phi/2f_\phi) = \sqrt{2 \lambda \Omega_\phi} \cos(\theta/2)/y_1$,  which shows the combined effect of all dynamical variables to get that of the axion field.}
\begin{equation}
\Omega^{1/2}_\phi \sin(\theta/2)  \equiv  \frac{\kappa \dot{\phi}}{\sqrt{6} H} \, , \quad \Omega^{1/2}_\phi \cos(\theta/2)
  \equiv \frac{\sqrt{2} \kappa m_\phi f_\phi}{\sqrt{3} H} \cos(\phi/2f_\phi) \, ,  \, y_1  \equiv \frac{2m_\phi}{H} \sin(\phi/2f_\phi) \, . \label{eq:defs}
\end{equation}
with which the KG equation can be written as a dynamical system in the form:
\begin{subequations}
\label{eq:new4}
  \begin{eqnarray}
  \theta^\prime &=& -3 \sin \theta + y_1 \, , \label{eq:new4a} \\
  y^\prime_1 &=& \frac{3}{2}\left( 1 + w_{tot} \right) y_1 + \frac{\lambda}{2}\Omega_{\phi}\sin \theta \,
  , \label{eq:new4b} \\
  \Omega^\prime_\phi &=& 3 (w_{tot} - w_\phi)
  \Omega_\phi \label{eq:new4c} \, .
\end{eqnarray}
\end{subequations}

Here, a prime denotes derivative with respect to the number of $e$-foldings $N \equiv \ln (a/a_i)$, with $a$ the scale factor of the Universe and $a_i$ its initial value. The decay constant appears explicitly in the newly defined (dimensionless) parameter $\lambda = 3/\kappa^2 f^2_\phi$, and then the FDM case with $\lambda =0$ (studied in Ref.~\cite{Urena-Lopez:2015odd}) is obtained in the limit $f_\phi \to \infty$. In contrast, we see that the mass parameter $m_\phi$ does not appear at all in the new equations of motion. Following the classification suggested in~\cite{Roy:2018nce}, the decay constant is an active parameter, whereas the mass is a passive one that does not have any influence in the evolution of the field $\phi$. The equation of state (EoS) for the axion field is directly related to the dynamical variable $\theta$ as,
\begin{equation}
  \label{eq:6}
  w_\phi \equiv \frac{p_\phi}{\rho_\phi} = \frac{x^2 - y^2}{x^2 +
    y^2} = - \cos \theta \, .
\end{equation}

Equations~\eqref{eq:new4} are a compact representation of the KG equation, and they reveal that the true variables driving the scalar field dynamics are $\left\lbrace \theta, y_1, \Omega_\phi \right\rbrace$. They also show that the effect of the trigonometric potential of Eq.~\eqref{eq:0} is encoded in one free parameter given by $\lambda$, and then it is possible to analyze the cosmological properties of both the axion field ($\lambda > 0$) and the FDM ($\lambda =0$) cases (see~\cite{Urena-Lopez:2015odd,Cedeno:2017sou}).

\subsection{Initial conditions}
For a correct numerical implementation of the equations of motion~\eqref{eq:new4} within a cosmological setting, it is necessary to estimate the right initial conditions of the dynamical variables at very early times. As done in Ref.~\cite{Urena-Lopez:2015odd} for the FDM case, in this section we find semi-analytical solutions for the radiation dominated era and extrapolate them to the present time. 

Assuming that all quantities are small and positive, i.e. $(\theta, y_1, \Omega_\phi) \ll 1$, Eq.~\eqref{eq:new4} takes the form (at linear order),
\begin{equation}
  \theta^\prime \simeq -3 \theta + y_1 \, , \quad y^\prime_1 \simeq 2
  y_1 \, , \quad \Omega^\prime_\phi \simeq 4 \Omega_\phi \,
  , \label{eq:7}
\end{equation}
whose analytical solutions are
\begin{equation}
   \theta = (1/5) y_1 + C (a/a_i)^{-3} \, , \quad y_1 = y_{1i} (a/a_i)^2 \, , \quad \Omega_\phi = \Omega_{\phi i} (a/a_i)^4 \, , \label{eq:8}
\end{equation}
where a subscript $i$ denotes the corresponding initial value for each variable. The solutions~\eqref{eq:8} are the same as those of the quadratic potential studied in~\cite{Urena-Lopez:2015odd}, basically because the second term on the rhs of Eq.~\eqref{eq:new4b} is of second order, which means that at early times the influence of $\lambda$ in the solutions is negligible. 

We now find a next-to-leading order solution for the initial conditions that takes into account the presence of $\lambda$, and for that we follow an iterative method. Let us consider the first order solutions~\eqref{eq:8}, substitute them in Eq.~\eqref{eq:new4b} and solve for a new solution of $y_1$. We find that
\begin{equation}
y_1 = 5 \theta_i (a/a_i)^2 + \frac{\lambda}{8} \Omega_{\phi i} \theta_i (a/a_i)^6 \, .
\label{y1fo}
\end{equation}
If we now use the foregoing solution and plug it into the right hand side of Eq.~\eqref{eq:new4a}, we find that a corrected solution for $\theta$ is
\begin{equation}
\theta = \theta_i (a/a_i)^2 \left[ 1 - \frac{\lambda}{72} \Omega_{\phi i} + \frac{\lambda}{72} \Omega_{\phi i} (a/a_i)^4 \right] \, , \label{thetafo}
 \end{equation}
whereas the solution for $\Omega_\phi$ remains the same. Let us assume that the axion field starts to behave as CDM at $a= a_{\rm osc}$, when it also starts to oscillate rapidly around the minimum of the potential and the EoS first passes through the value $w_\phi =0$ (corresponding to $\theta = \pi/2$). From the combination of the above equations, we obtain from the matching condition at $a= a_{\rm osc}$ that
\begin{subequations}
\label{eq:initials}
\begin{equation}
a^2_{\rm osc} \left( 1 + \frac{\lambda}{72} \frac{\Omega_{\phi 0}}{\Omega_{r 0}} a_{\rm osc} \right)  = \frac{\pi \, \theta^{-1}_i a^2_i}{2\sqrt{1+\pi^2/36}} \, .
\label{cqe}
\end{equation}
Notice that for $\lambda =0$ we recover, as expected, the required matching equation for the quadratic potential (see~\cite{Urena-Lopez:2015odd} for more details). As shown in the appendix~\ref{sec:higher-order}, the iterative integration method could be used again to generate a higher-order equation to determine $a_{\rm osc}$, but we will restrict ourselves to Eq.~\eqref{cqe} as it is enough for the purposes of this paper.

In contrast to the FDM case, there is an additional trigonometric constraint that is characteristic of the axion potential, and that can be obtained directly from the definitions~\eqref{eq:defs},
\begin{equation}
4 \frac{m^2_\phi}{H^2_i} = y^2_{1i} + 2 \lambda \Omega_{\phi i} \, . \label{eq:trigonometric}
\end{equation}
Although we use it only as an additional constraint for the initial conditions, it should be emphasized that Eq.~\eqref{eq:trigonometric} is of general applicability at all times. Again, for the case $\lambda =0$ we recover the usual expression of the FDM case, namely $y_{1i} = 2m_{\phi}/H_i$. Hence, the initial conditions in the general case are obtained from the combined solution of Eqs.~\eqref{cqe},~\eqref{eq:trigonometric} and
\begin{equation}
y_{1i} = 5\theta_i \left( 1 + \frac{\lambda}{40} \Omega_{\phi i} \right)  \, , \quad \Omega_{\phi i} = \frac{a^4_i}{a^3_{\rm osc}} \frac{\Omega_{\phi 0}}{\Omega_{r 0}} \, .
\end{equation}
\end{subequations}

The initial conditions~\eqref{eq:initials} are further adjusted by means of the shooting procedure implemented in \texttt{CLASS} to give the right current values of the physical parameters. The values of $a_{\rm osc}$ for different values of $\lambda$, as obtained from the numerical solutions, are shown in table~\ref{table_1}, where it can be seen that the onset of the scalar field oscillations suffers a delay as $\lambda$ increases. Because of numerical limitations, there is a maximum value of $\lambda$ for each one of $m_{\phi}$ that we can consider. For larger values of $\lambda$, it is difficult to calculate the initial conditions because of the exponential sensitivity that appears in the estimation of $a_{\rm osc}$, see Appendixes~\ref{sec:higher-order} and~\ref{sec:ewdm} for more details. 

\begin{table}[h!]
\centering
\begin{tabular}{|c|c|c|c|c|c|c|}
\hline
$\lambda$ & 0 & 10 & $10^2$ & $10^3$ & $10^4$ & $10^5$ \\ \hline 
$\log(a_{\rm{osc}})$ & -6.159 & -6.159 & -6.159 & -6.143 & -6.048 & -5.838 \\ \hline 
$\delta \theta_0$ & --- & $174^\circ$ & $162^\circ$ & $124^\circ$ & $44^\circ$ & $0.47^\circ$ \\ \hline 
\end{tabular}
\caption{\label{tab:table1} Numerical values for the onset of oscillations of the axion field for each value of $\lambda$ and fixed axion mass $m_\phi=10^{-22}$eV. For $\lambda = 0, 10, 10^2$, oscillations start at the same time, whereas for $\lambda = 10^3, 10^4, 10^5$, we notice that oscillations start later as $\lambda$ increases. In the last row we show the initial field displacement from the top of the axion potential, for a comparison with the Extreme axion Wave Dark Matter model~\cite{Zhang:2017flu}, see Appendix~\ref{sec:ewdm} for details.}
\label{table_1}
\end{table}

\begin{center}
\begin{figure}[htp!]
    \centering
   \includegraphics[width=0.6\linewidth]{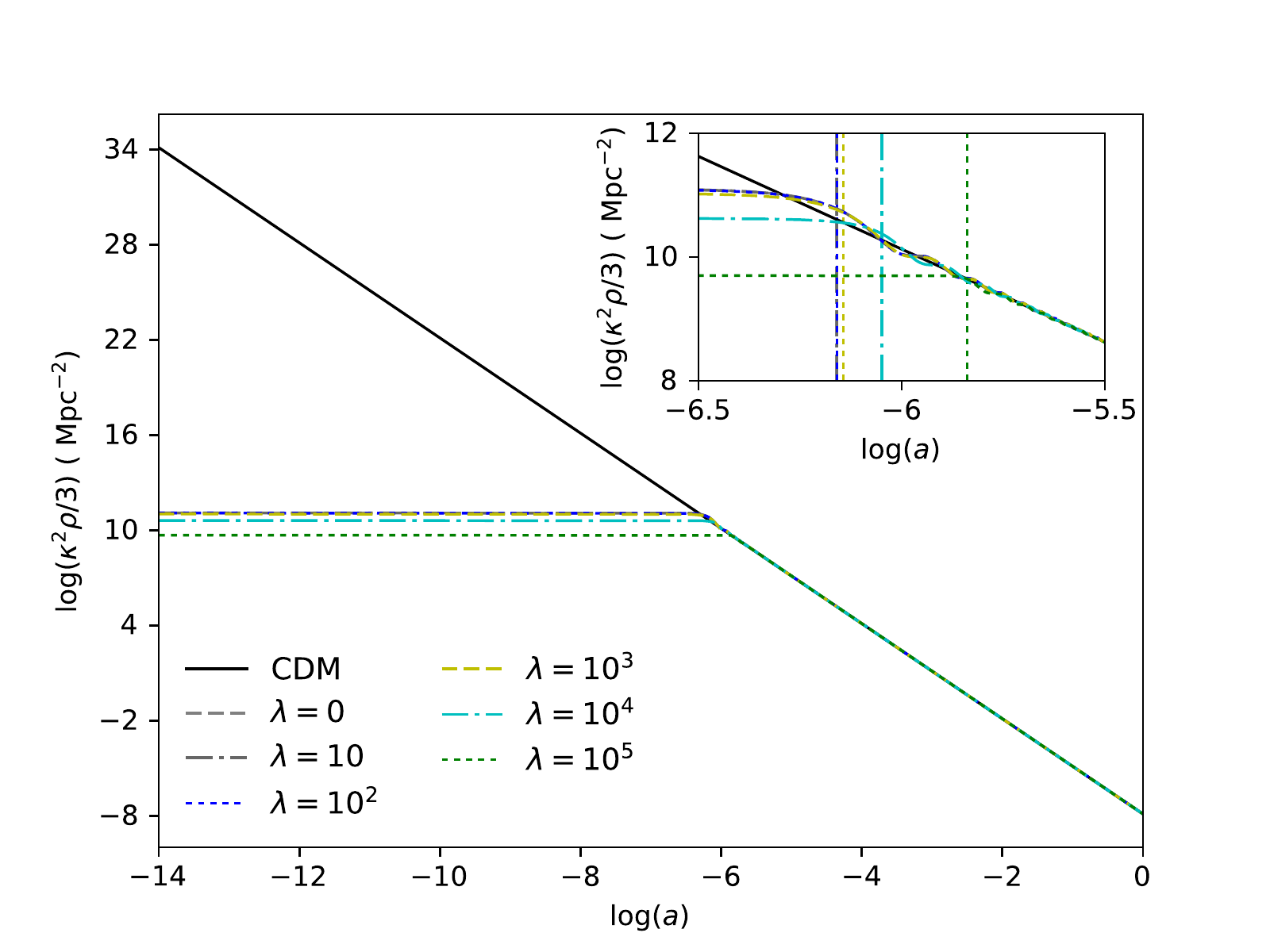}
        \label{fig:first_sub}   
    \caption{Evolution of CDM and SFDM energy density for a fixed axion mass of $m_\phi=10^{-22}$eV, and different values of the decay parameter $\lambda$ of the potential \eqref{eq:0}. Initially the amplitude of the axion energy density is less than that of the CDM, but once the axion field starts to oscillate (around $a=10^{-6}$), it evolves just as the CDM case. Inset: It can be noticed that larger values of parameter $\lambda$ delay the scalar field oscillations, then the axion field evolves as CDM. Vertical lines indicate the onset of oscillations $\log(a_{\rm{osc}})$ for each value of $\lambda$, see also table~\ref{table_1}.}
    \label{fig:sample_subfigures}
    \label{plot1}
\end{figure}
\end{center}

A comparison of the evolution of the CDM and axion densities is shown in Figure~\ref{plot1}. While the CDM density redshifts as $a^{-3}$, we see that the axion energy density remains constant before the start of the field oscillations at $a=a_{\rm osc} \simeq 10^{-6}$, but afterwards the two densities evolve together. As also shown in the inset, the onset of the field oscillations depends on the value of the decay constant through the parameter $\lambda$, and in general the oscillations are delayed as the value of $\lambda$ increases, which is consistent with the numerical results shown in table~\ref{table_1}. We can also notice that the transition of the axion energy density to the CDM behavior happens more abruptly for larger values of $\lambda$, which is one of the consequences of the aforementioned exponential sensitivity of the numerical solutions on the initial conditions that we discussed above.

\subsection{Linear Perturbations}
Now, we consider linear perturbations around the background values of the FRW line element (in the synchronous gauge) as well as for the scalar field in the following form:
\begin{equation}
ds^2 = -dt^2+a^2(t)(\delta_{ij}+h_{ij})dx^idx^j\, ,\ \phi(x,t) = \phi(t)+\varphi(x,t)\, ,
\label{mfpert}
\end{equation}
where $h_{ij}$ and $\varphi$ are the metric and scalar field perturbations respectively. The linearized KG equation, in Fourier space and for a general potential, reads \cite{Ratra:1990me,Ferreira:1997au,Ferreira:1997hj,Perrotta:1998vf}:
\begin{equation}
  \ddot{\varphi} = - 3H \dot{\varphi} - \left(\frac{k^2}{a^2} + \frac{\partial^2 V(\phi)}{\partial \phi^2}\right) \varphi -
  \frac{1}{2} \dot{\phi} \dot{\bar{h}} \, , \label{eq:13}
\end{equation}
where $\bar{h} = {\bar{h}^j}_j$ is the trace of scalar metric perturbations, and $k$ is the comoving wavenumber. Although a functional dependence of the scalar field perturbation is not explicitly shown, note that Eq.~\eqref{eq:13} is written for a Fourier mode
$\varphi(k,t)$. After a change of variables to the new quantities $\delta_0$ and $\delta_1$ (see Appendix~\ref{sfp} for details), Eq.\eqref{eq:13} is described by the following system of first order differential equations~\cite{Cedeno:2017sou},
\begin{subequations}
\label{eqdeltas}
\begin{eqnarray}
\delta^\prime _0 &=&  \left[-3\sin\theta-\frac{k^2}{k_J^2}(1 - \cos \theta) \right]\delta_1+\frac{k^2}{k_J^2}\sin \theta \delta_0 - \frac{\bar{h}'}{2}(1-\cos\theta)\, , \label{d0p} \\
\delta^\prime_1 &=& \left[-3\cos \theta- \left( \frac{k^2}{k_J^2} - \frac{\lambda \Omega_\phi}{2y_1} \right) \sin \theta \right] \delta_1 + \left( \frac{k^2}{k_J^2} - \frac{\lambda \Omega_\phi}{2y_1}\right) (1 + \cos \theta) \, \delta_0 - \frac{\bar{h}'}{2} \sin \theta \, ,
\end{eqnarray}
\end{subequations}
where we defined the (squared) Jeans wavenumber as $k_J^2=H^2a^2y_1$, which is the same definition used for the case of a quadratic potential~\cite{Urena-Lopez:2015odd}.

The density and pressure contrasts $\delta_\phi$, $\delta p_\phi$, and velocity divergence $\theta_\phi$, are given by the standard definitions~\cite{Perrotta:1998vf,Hu:1998kj,Ferreira:1997hj}, and in terms of the new perturbation variables they take the form:
\begin{subequations}
   \label{eq:26}
  \begin{eqnarray}
    \delta_\phi = \frac{\dot{\phi}  \dot{\varphi} + \partial_\phi V \,
    \varphi}{\dot{\phi}^2/2 + V(\phi)} = \delta_0\, , \quad \delta_{p_\phi} = \frac{\dot{\phi}  \dot{\varphi} - \partial_\phi V \,
    \varphi}{\dot{\phi}^2/2 + V(\phi)} = \sin \theta \delta_1 - \cos\theta \delta_0 \, , \label{eq:26a} \\
    (\rho_\phi + p_\phi) \theta_\phi = (k^2/a) \dot{\phi} \varphi = \frac{k^2}{2am_\phi}\rho_\phi \left[ \left(1 - \cos \theta \right)\delta_1 - \sin\theta\delta_0 \right] \, . \label{eq:26c}
  \end{eqnarray}
\end{subequations}
It is important to mention that we have gained physical interpretation for the new dynamical variable $\delta_0$: it plays the role of the scalar field density contrast,  $\delta_\phi$, according to the first expression in Eq.~\eqref{eq:26a}. This implies that Eq.~\eqref{d0p} is the closest we can get of a fluid equation for the scalar field perturbations. The interpretation of $\delta_1$ remains elusive, and it remind us of the difficulties to match Eq.~\eqref{eq:13} to a fluid even in the generalized case~\cite{Hu:1998kj} (although see Appendix~\ref{sec:eawdm}).

For the particular case of the axion field endowed with the potential~\eqref{eq:0}, the expressions~\eqref{eqdeltas} now reads
\begin{subequations}
\label{eqnewdeltas}
\begin{eqnarray}
\delta^\prime_0 &=&  \left[-3\sin\theta-\frac{k^2}{k^2_J}(1 - \cos \theta) \right] \delta_1 + \frac{k^2}{k^2_J} \sin \theta \delta_0  - \frac{\bar{h}^\prime}{2}(1-\cos\theta) \, , \label{d0} \\
\delta^\prime_1 &=& \left[-3\cos \theta - \frac{k_{eff}^2}{k_J^2} \sin\theta \right] \delta_1 + \frac{k_{eff}^2}{k_J^2} \left(1 + \cos \theta \right) \, \delta_0 - \frac{\bar{h}^\prime}{2} \sin \theta \, ,  
\label{d1}
\end{eqnarray}
\end{subequations}
where we have defined an effective wavenumber of the perturbations as $k^2_{eff} \equiv k^2 - \lambda a^2 H^2 \Omega_\phi/2$. 

The equations of linear perturbations for the standard FDM case are again obtained when $\lambda =0$, for which $k^2_{eff} = k^2$ is just the standard Laplacian term in Fourier space. Because now $y_1 = 2m_\phi/H$, the Jeans wavenumber $k_J$ is then the only characteristic scale in the evolution of linear perturbations, and the responsible for the appearance of a sharp cut-off in their mass power spectrum: linear perturbations are heavily suppressed for wavenumbers $k > k_J$. The Jeans wavenumber is always proportional to the geometric mean of the Hubble parameter $H$ and the boson mass $m_\phi$, namely $k_J = a \sqrt{2H m_\phi}$, which shows that the cut-off in the MPS is sensitive to both the parameters of the axion model and to the background expansion. More details about the cut-off of linear perturbations in the FDM case $\lambda =0$ can be found in~\cite{Urena-Lopez:2015odd}.

\subsection{Tachyonic instability}\label{ti}
One of the main effects on linear perturbations of axion fields (for $\lambda > 0$) is the appearance of an enhancement in the growth of the density contrast $\delta_0$, that was first discussed in~\cite{Cedeno:2017sou,Zhang:2017chj,Zhang:2017flu} and thereby dubbed as a \textit{tachyonic instability} because $k^2_{eff} < 0$. Such instability provokes the appearance of a bump in the MPS of the perturbations that is well localized in wavenumbers around the Jeans one $k_J$.

To have a qualitative understanding of the tachyonic instability, we follow and extend the procedure already outlined in~\cite{Cedeno:2017sou}. Let us write Eqs.~\eqref{eqnewdeltas} on  rapid oscillations regime, under which all trigonometric terms are time-averaged to zero, $\langle \sin \theta \rangle = \langle \cos \theta \rangle = 0$.\footnote{Recently, the authors in~\cite{Cookmeyer:2019rna} made a comparison of the different approximations one can find in the literature to follow the cosmological evolution of ultra-light bosons. Such approximations, which correspond to diverse choices in cycle-averaging procedures, are necessary to deal with the rapid oscillations of the axion field at late times, see the original field equations~\eqref{eq:2} and~\eqref{eq:13}. It was there concluded that our approximation method, which has been used previously in Refs~\cite{Urena-Lopez:2015gur,Cedeno:2017sou}, is the closest, compared to others, to the exact solution of the field equations of motion.} Hence, we find
\begin{subequations}
\begin{equation}
\delta^\prime_0 =  -\frac{k^2}{k^2_J} \delta_1 - \frac{\bar{h}^\prime}{2} \, ,\quad \delta^\prime_1 = \frac{k_{eff}^2}{k_J^2} \delta_0 \, .  
\label{eqnewdeltas1}
\end{equation}
If we neglect, for simplicity, the time variation of both $k_J$ and $k_{eff}$, the foregoing equations can be combined into the form of a forced harmonic oscillator for the density contrast, namely,
\begin{equation}
\delta^{\prime \prime}_0 + \omega^2 \delta_0 = - \frac{\bar{h}^{\prime \prime}}{2} \, , \quad \omega^2 \equiv \frac{k^2 k^2_{eff}}{k^4_J} \, . \label{eq:harmonic}
\end{equation}
\end{subequations}
From the above we see that the tachyonic instability requires of two conditions. Firstly, the start of rapid oscillations of the field around the minimum of its potential, and secondly, a negative squared amplitude of the angular frequency, $\omega^2 < 0$, in Eq.~\eqref{eq:harmonic}. The latter condition is possible because the effective wavenumber $k^2_{eff}$ can be either positive or negative, although it depends on a non-simple combination of the cosmological quantities $a$, $H$ and $\Omega_\phi$. 

\begin{center}
\begin{figure}[htp!]
    \centering
            \centering
        \begin{subfigure}[b]{0.45\textwidth}  
            \centering 
            \includegraphics[width=\textwidth]{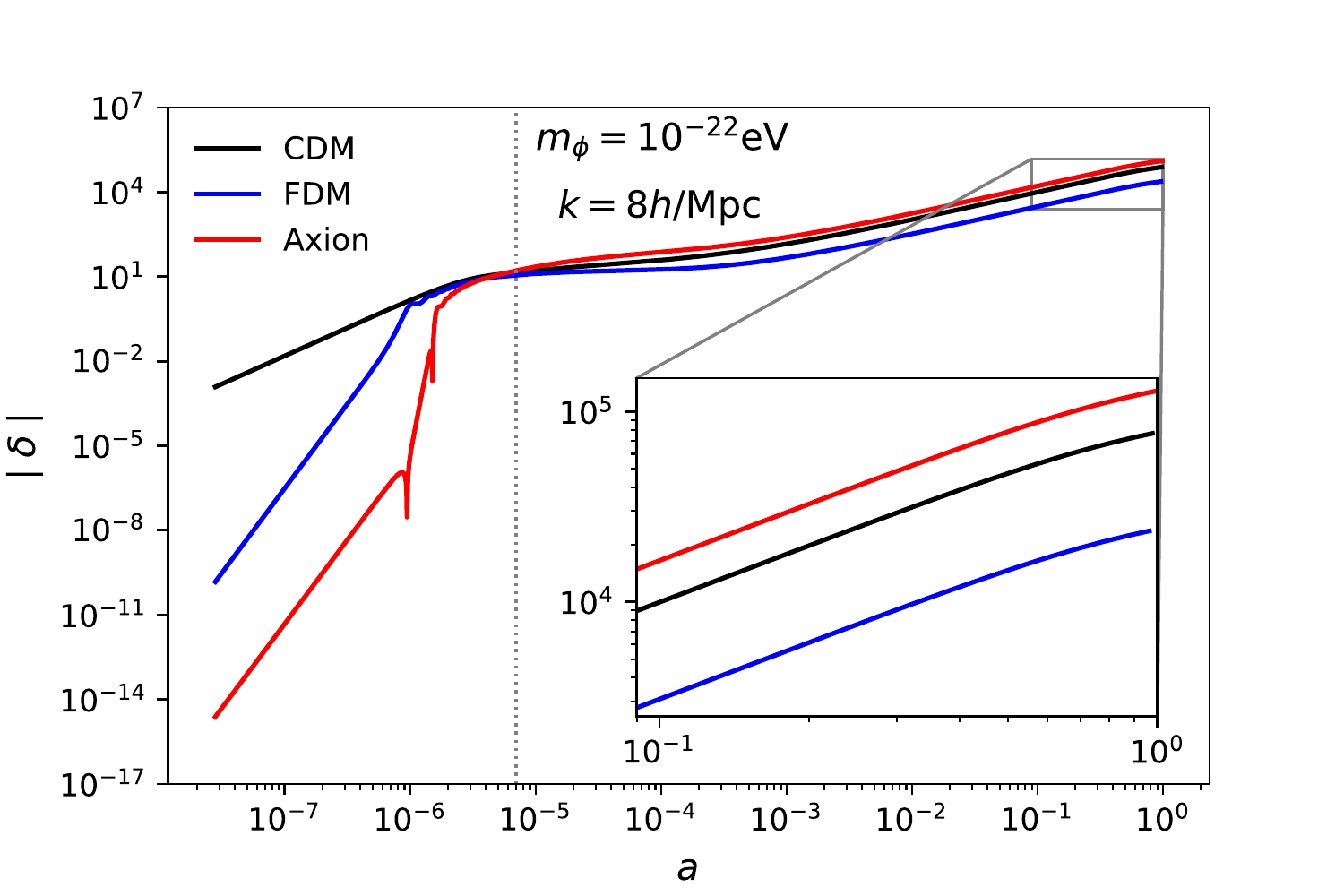}
        \end{subfigure}
        \begin{subfigure}[b]{0.48\textwidth}   
            \centering 
            \includegraphics[width=\textwidth]{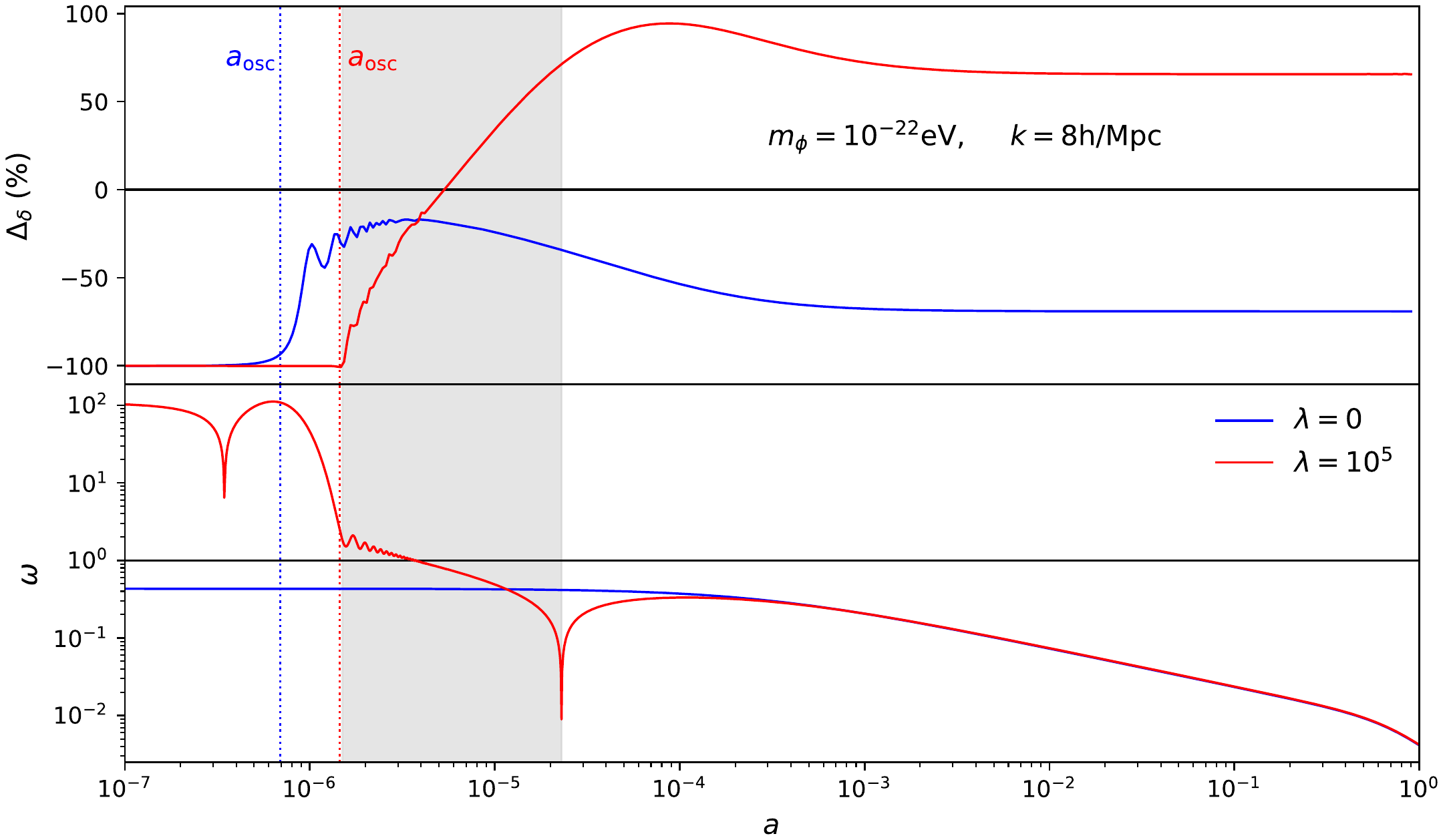}
        \end{subfigure}
    \caption{(Left) Cosmological evolution of the CDM density contrast (black), FDM ($\lambda=0$, blue) and the axion field ($\lambda=10^5$, red) with $m_{\phi}=10^{-22}$eV and $k=8h/$Mpc. The vertical gray dotted line indicates the time from which the density contrast for the FDM (axion) evolves at a constant rate with respect to CDM but with less (more) amplitude. (Right) Evolution of the relative difference between density contrasts $ \Delta_\delta$ (top) and frequency $|\omega|$ (bottom) for a fixed axion mass of $m=10^{-22}$eV and wavenumber $k = 8h/\rm{Mpc}$, evaluated at $\lambda = 0$ (solid blue line) and $\lambda = 10^5$ (solid red line). The light-gray region indicates the duration of the tachyonic effect on the density contrast, while the vertical dotted blue and red lines show the onset of the oscillation of the scalar field for $\lambda = 0$ and $\lambda = 10^5$ respectively. See text for more details.}
    \label{dcomega1}
\end{figure}
\end{center}

\begin{center}
\begin{figure}[htp!]
    \centering 
    \includegraphics[width=0.7\textwidth]{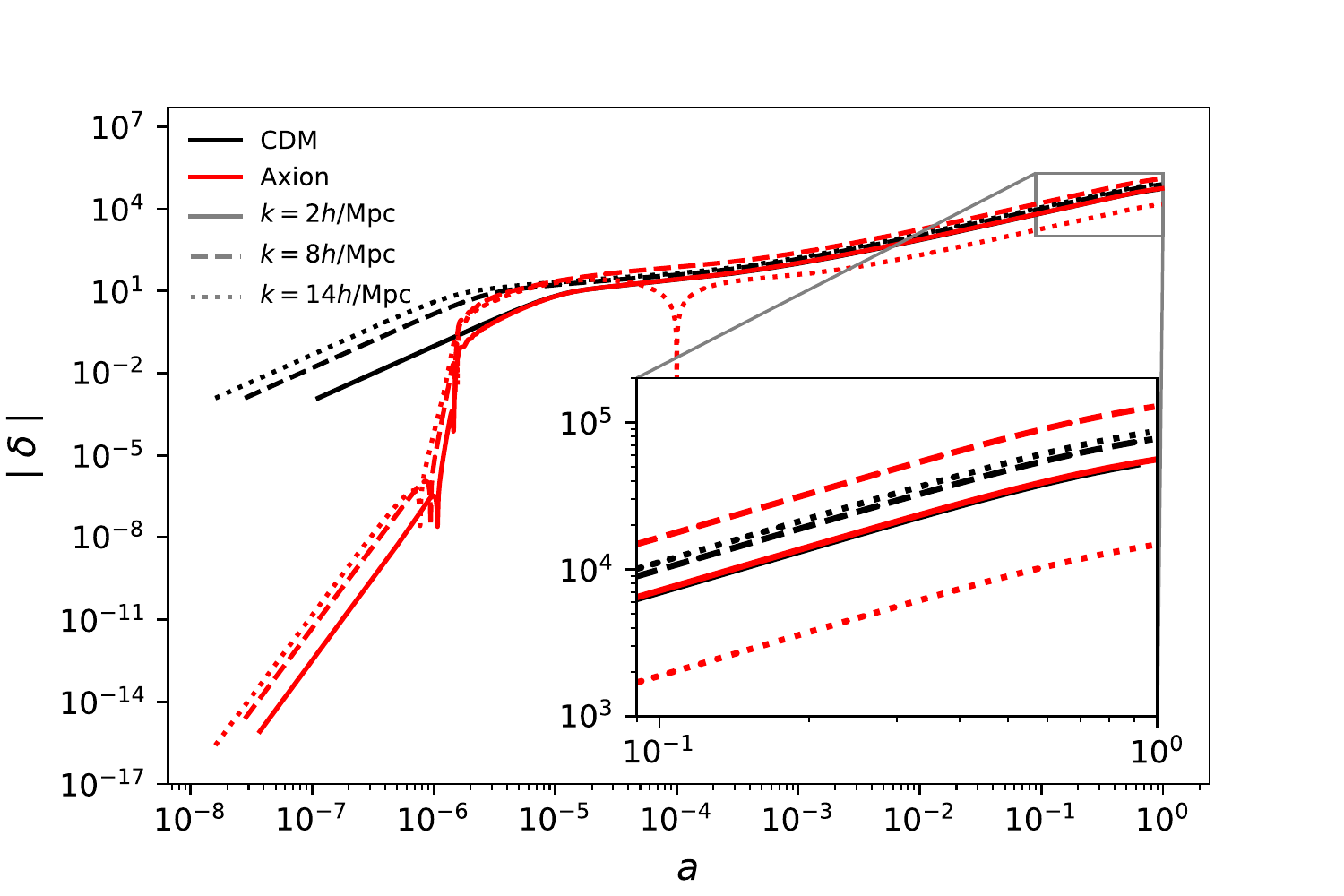}
    \caption{The amplitude of the density contrast for CDM (black) and axion field (red) with $m_\phi = 10^{-22}$eV and $\lambda = 10^5$ for three different modes: $k=2h/$Mpc (solid curve), $k=8h/$Mpc (dashed curve), and $k=14h/$Mpc (dotted curve). Each mode has a specific tachyonic instability duration, and the differences with respect to CDM due to such effect can be appreciated. See the text for more details.}
    \label{dcomega2}
\end{figure}
\end{center}

The tachyonic instability and the conditions for its appearance are illustrated in Figures~\ref{dcomega1} and~\ref{dcomega2}. The left panel of Figure~\ref{dcomega1} shows the cosmological evolution of the density contrast $\delta_{\phi}$ for FDM ($\lambda=0$) and axion field ($\lambda=10^5$), with $m_{\phi}=10^{-22}$eV, for the wavenumber $k=8h/$Mpc. The corresponding case for CDM is also shown for comparison. A similar analysis is presented on the upper plot of the right panel of Figure~\ref{dcomega1}, but now in terms of the relative difference between the axion density contrast with respect to CDM: $\Delta_\delta \equiv (\delta_\phi - \delta_{CDM} )/\delta_{CDM}$, whereas the lower plot shows the evolution of the angular frequency $|\omega|$ defined in Eq.~\eqref{eq:harmonic}. The light-gray region indicates the period of time when the tachyonic condition is acting on the density model in the case $\lambda =10^5$, starting from the onset of the scalar field oscillations together with the condition $\omega^2 < 0$.

In the FDM case, the density contrast $\delta_{\phi}$ for the chosen wavenumber can not catch up completely with the CDM solution after the onset of rapid oscillations (at around $a \simeq 10^{-6.16}$), but nonetheless keeps a constant ratio with respect to CDM at late times (this constant ratio can be explained in terms of the growth factor, see Sec.~\ref{sec:observables} below). The particular value $k=8 \, h/\mathrm{Mpc}$ was chosen because it corresponds, approximately, to the cut-off scale for $m_{\phi} = 10^{-22}$eV. In contrast, for the value $\lambda = 10^5$ we see that the onset of oscillations is delayed (at $a_{{\rm{osc}}} = 10^{-5.84}$), but the amplitude of the axion density contrast $\delta_0$ grows quickly reaching larger values than that of CDM. This growth persists until just after the tachyonic instability disappears (when once again $\omega^2 > 0$) at around $a \simeq 10^{-4.8}$. After this, the density contrast $\delta_0$ then evolves like in the FDM case and keeps a constant amplitude with respect that of CDM at late times. As we will see in Sec.~\ref{sec:observables}, the tachyonic instability in the case of axion density contrast will manifest in the MPS as a bump at large wavenumbers, and the maximum of such bump will be located precisely at $k\simeq 8h/$Mpc.

A qualitative understanding of the growth of density perturbations in both the FDM and axion cases can be obtained as follows (see also the analyses in~\cite{Urena-Lopez:2015gur,Cedeno:2017sou}). Let us write the angular frequency defined in Eq.~\eqref{eq:harmonic} in the form
\begin{equation}
    \omega^2 = \frac{k^2}{a^2 H^2 y^2_1} \left( \frac{k^2}{a^2 H^2} - \frac{\lambda}{2} \Omega_\phi \right) \, .
\end{equation}
In the FDM case ($\lambda$ =0) we see that the angular frequency is always positive, $0 < \omega^2$, and the tachyonic instability never happens. Moreover, modes that comply with the condition $\omega^2 >1$, that is $k > a H y^{1/2}_1$, are expected to have a suppressed growth with respect to the CDM case. Notice that after the onset of rapid oscillations $y_1 > 1$, and then these modes are within the Hubble horizon and satisfy the condition $k > a H$. The minimum wavenumber that suffers suppression is given by $k_{min} = a_{osc} H_{osc} y^{1/2}_{1,osc}$. In the FDM case $y_1 = 2 m_\phi/H$ and $H_{osc} \simeq m_\phi/3$, and then $k_{min} \simeq \sqrt{2/3} \, a_{osc} m_\phi = 1.27 \times 10^7 a_{osc} (m_\phi/10^{-22}\mathrm{eV}) \, \mathrm{Mpc}^{-1}$. In the fiducial case $m_\phi = 10^{-22} \, \mathrm{eV}$, for which $a_{osc} = 10^{-6.159}$ (see Table~\ref{tab:table1}), we find that modes for which $8 \, \mathrm{Mpc}^{-1} \lesssim k$ should appear suppressed with respect to the CDM case. Although the above estimation assumes that $\omega^2 = \mathrm{const.}$, the results are in agreement with the full numerical solutions in Sec.~\ref{sec:observables} below.

For the axion case ($\lambda > 0$), the above argument just needs some adjustments. First, notice that the evolution of the modes is the same as in the FDM case if both of the following conditions are satisfied: (i) $k/(aH) \gg y_1$ and (ii) $k/(aH) \gg \sqrt{\lambda \Omega_\phi/2}$. Condition (i) tells us that the mode should already be inside the horizon after the onset of rapid oscillations and that growth suppression should occur for $k \gg k_{min} = a_{osc} H_{osc} y_{1,osc}$. Given that the solution of $y_1$ is similar to the FDM case, $y_{1,osc} \simeq 2 m_\phi/H_{osc}$, see Eq.~\eqref{eq:new4b}, we obtain this time that $k_{min} = 2 a_{osc} m_\phi \simeq 3.12 \times 10^7 a_{osc} (m_\phi/10^{-22}\mathrm{eV}) \, \mathrm{Mpc}^{-1}$. For the values $m_\phi = 10^{-22} \, \mathrm{eV}$ and $\lambda = 10^5$, we find from Table~\ref{tab:table1} that $a_{osc} = 10^{-5.838}$, and then $k_{min} \simeq 45 \, \mathrm{Mpc}^{-1}$. The result is a shift to larger wavenumbers for the suppression to happen, as a direct consequence of the delayed start of the rapid oscillations as compared to the FDM case.

Condition (ii) is necessary to guarantee $\omega^2 > 0$, but it also tells us about the role of the extra parameter $\lambda$ in the recovery of the FDM behavior of the density modes. Given that $\Omega_\phi \leq \Omega_{\phi 0}/(\Omega_{\phi 0} +\Omega_{b 0}) \simeq \mathcal{O}(1)$, then conditions (i) and (ii) together imply that the FDM result is recovered for modes $k > \mathrm{max} (2,\sqrt{\lambda/2}) \times a_{osc} m_\phi$. The latter expression also indicates the range of modes that are subjected to the tachyonic instability, for which $\omega^2 < 0$, these are $2 \lesssim k/(a_{osc} m_\phi) < \sqrt{\lambda/2}$. This explains why one requires a large enough value of $\lambda >  8$ for the instability to appear in a noticeable range of wavenumbers, and also that the tachyonic enhancement should be most prominent for wavenumbers close to the lower limit $k \gtrsim k_{min}$, i.e. close to the cut-off scale of the FDM case. In this respect, it is the boson mass that determines the characteristic scales for structure formation for both the FDM and axion models. We must recall that the tachyonic growth of the density modes is a cumulative (integrated) effect over time, and then the differences in comparison to FDM are more noticeable for larger values of $\lambda$.

For a comparison of our qualitative analysis above with full numerical solutions, we show in Figure~\ref{dcomega2} the cosmological evolution of the density contrast for CDM (black curves) and the axion field (red curves) with $m_{\phi} = 10^{22}$eV and $\lambda=10^5$, considering three modes with different tachyonic instability duration. The mode $k=2h/$Mpc (solid curves) is almost unaffected by the tachyonic instability since for it $0 \lesssim \omega^2 < 1$, and its late time evolution is basically the same of CDM, $\delta_{\phi}\simeq \delta_{CDM}$. Then we have $k=8h/$Mpc (dashed curve), which is the most affected mode since for it $\omega^2 \lesssim -1$; it can be seen that at late times $\delta_{\phi}>\delta_{CDM}$. Finally, we show the case for $k=14h/$Mpc (dotted curves), which does not experiences the tachyonic instability regime as for it $\omega^2 \gtrsim 1$, and then at late times this mode appears as suppressed in comparison to CDM: $\delta_{\phi}<\delta_{CDM}$, as can be observed in the plot.

Summarizing, we find that for large scales $k \lesssim 2 h/\rm{Mpc}$ the tachyonic instability is practically non-existent, and also for them the condition $0 < \omega^2 \ll 1$ is accomplished at all times. The evolution of the density contrast for these scales is governed by the equation $\delta^{\prime \prime}_0 \simeq -(1/2) \bar{h}^{\prime \prime}$ (see Eq.~\eqref{eq:harmonic}), and we obtain for them the same solution as for CDM linear perturbations, that is $\delta_0 \simeq -(1/2) \bar{h}$. Likewise, small scales $k \gtrsim 22 h/\rm{Mpc}$ are also always free from tachyonic instabilities as for them $\omega^2 \gtrsim 1$ at all times. The latter condition means that they do not longer grow with the CDM solution, but now they must be suppressed as in the standard FDM case. Therefore, wavenumbers within the range $2 \lesssim k/ (h/\rm{Mpc}) \lesssim 22$ will present an enhancement in their density contrast amplitude, which is again in agreement with our qualitative analysis above.

\section{Cosmological constraints}\label{sec:observables}
The solutions of Eq.~\eqref{eqnewdeltas} are useful to build up cosmological observables such as the CMB anisotropies and the MPS, which can then be contrasted with observations. In this section we first present a qualitative comparison with the observables, and then present the details and results from a  parameter estimation procedure. 

\subsection{CMB anisotropies}\label{cmb_anisot}
\begin{figure*}[htp!]
        \centering
        \begin{subfigure}[b]{0.49\textwidth}  
            \centering 
            \includegraphics[width=\textwidth]{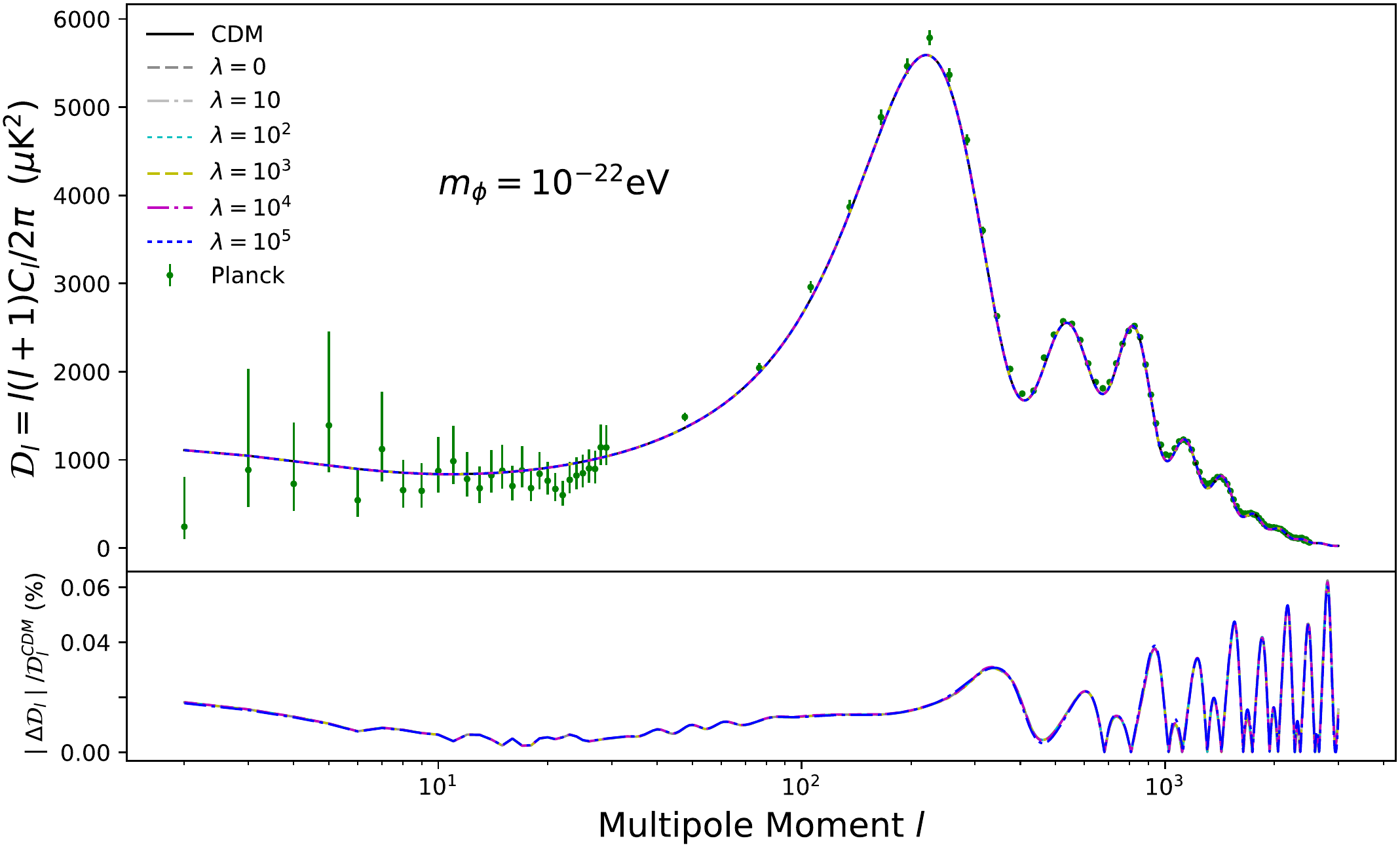}
            \label{fig:mean and std of net24}
        \end{subfigure}
        \begin{subfigure}[b]{0.49\textwidth}   
            \centering 
            \includegraphics[width=\textwidth]{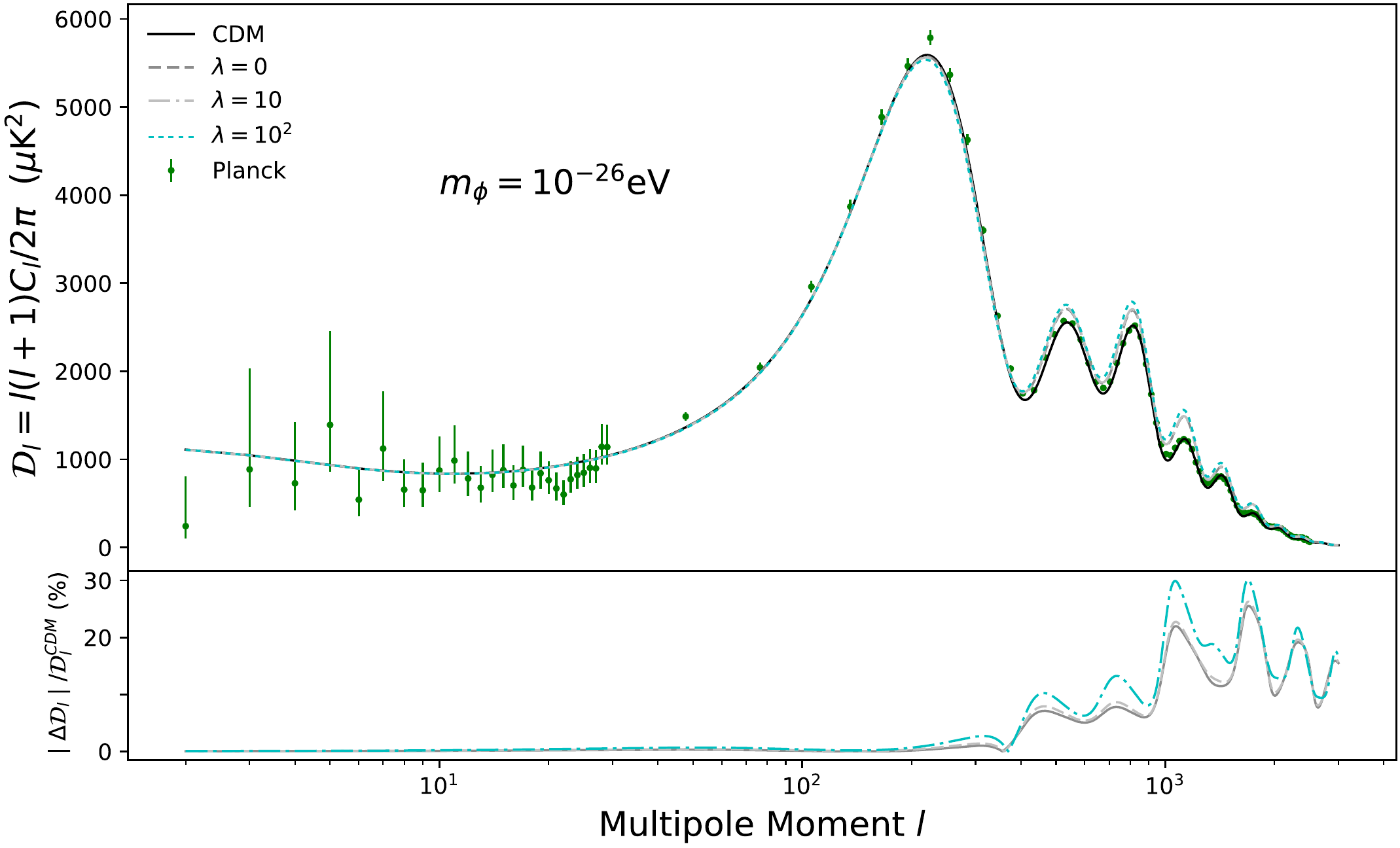}
            \label{fig:mean and std of net44}
        \end{subfigure}
        \caption[ The average and standard deviation of critical parameters ]
        {Temperature Power Spectrum for CDM and SFDM for two axion masses: $m_\phi=10^{-22},10^{-26}$eV. The effect of $\lambda$ is clearly noted for the latter where, for large multipoles, the differences are greater as the value of $\lambda$ increases. See text for more details.} 
        \label{tps}
    \end{figure*}

The CMB power spectrum for both CDM and SFDM, for a couple of values of the axion mass, is shown in Figure~\ref{tps}, where we have included data from the Planck Collaboration.\footnote{Based on observations obtained with Planck (http://www.esa.int/Planck), an ESA science mission with instruments and contributions directly funded by ESA Member States, NASA, and Canada.}
For a fiducial axion mass of $m_\phi = 10^{-22}$eV (left panel) we observe that, regardless of the value of $\lambda$, the axion field reproduces the CMB spectrum as good as CDM. In fact, the major discrepancy between both cases is of $\sim 0.06\%$ for large multipoles. In contrast, for an axion mass of $m_\phi = 10^{-26}$eV (right panel), we clearly note that the CMB spectrum does not fit the observational data, with a major discrepancy of $\sim 30\%$ for $l\sim 10^3$. 

\begin{figure}[h!]
    \centering
   \includegraphics[width=0.7\textwidth]{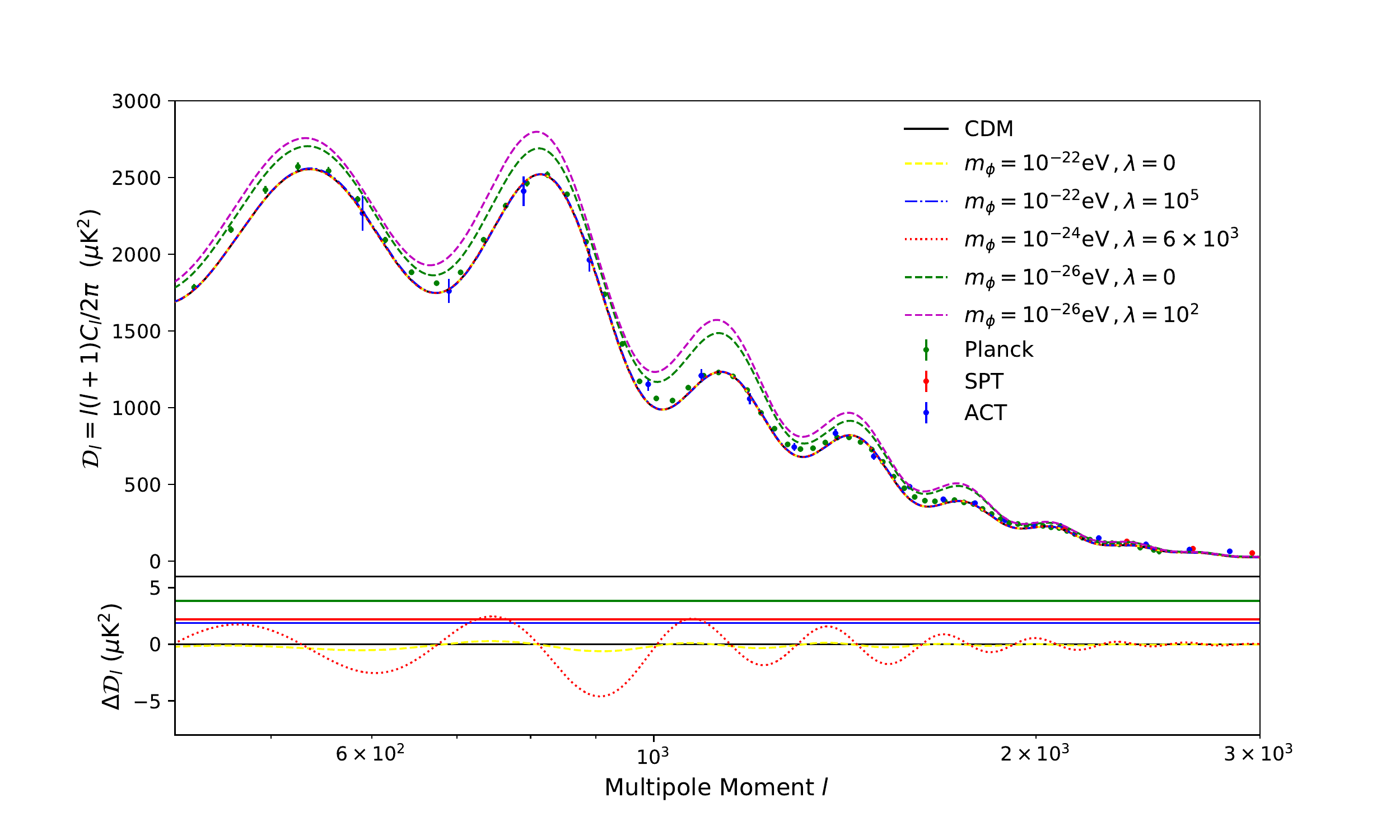}
        \label{fig:first_sub}   
    \caption{CMB anisotropies for high multipoles. Data from Planck (green dots), SPT \cite{George:2014oba} (red dots) and  ACT \cite{Das:2013zf} (blue dots) are shown to compare with our numerical solutions. For masses lighter as $m_\phi = 10^{-26}$eV and $\lambda = 0, 10^2$, we found notorious discrepancies with the observational data. The lower panel shows the relative differences between CDM and SFDM with $\left\lbrace m_\phi=10^{-22}\rm{eV},\lambda=0 \right\rbrace$ (yellow dashed line) and $\left\lbrace m_\phi=10^{-24}\rm{eV},\lambda=6\times 10^3 \right\rbrace$ (red dotted line). The horizontal green, red and blue lines indicate the minimum sensitivity for Planck, SPT and ACT observations respectively, given by $(\sigma_{\text{Planck}},\sigma_{\text{SPT}},\sigma_{\text{ACT}}) = (3.8,2.2,1.9)\mu$K$^2$. }
    \label{highl}
\end{figure}

We have also considered CMB observations for high multipoles, as can be seen in Figure~\ref{highl}, where we have also included a wider range of values for $m_\phi$ and $\lambda$. We observe in the upper panel that for large multipoles the case of an axion mass of $10^{-26}$eV with quadratic ($\lambda=0$) and trigonometric potential ($\lambda=10^2$) still have more amplitude than the extreme case with $m_\phi = 10^{-22}$eV and $\lambda=10^5$. In particular, we observe that for a given axion mass the effect of consider $\lambda > 0$ is to increase the amplitude of the CMB power spectrum in comparison with the FDM case (this can be clearly seen when $m_{\phi} = 10^{-26}$eV). The lower panel shows that observations from Planck satellite, the South Pole Telescope (SPT), and the Atacama Cosmology Telescope (ACT), do not constrain the fiducial case of a free axion with mass $10^{-22}$eV and $\lambda = 0$. However, we note that ACT rules out a combination of parameters such as $m_{\phi}=10^{-24}$eV and $\lambda=6\times 10^3$. As was previously noted in~\cite{Hlozek:2014lca}\footnote{For such light axion masses to be consistent with CMB anisotropy data, less amount of SFDM is required: if $\Omega_{m} = \Omega_{\phi} + \Omega_{CDM}$, then $\Omega_{\phi}/\Omega_m<0.05$ and $\Omega_{\phi}h^2\leq 0.006$ at 95\% confidence~\cite{Hlozek:2014lca}. In our analysis, we are considering that the axion field provides all the DM in the universe, and then CMB observations are able to impose constraints on lighter masses. This will be discussed later in Section 3.4.}, lighter scalar fields leave more noticeable effects on the acoustic peaks  because their equation of state keeps the value $w_{\phi}=-1$ for a longer time, during radiation domination era, than heavier ones. Our results show that, for the same mass, a value $\lambda \neq 0$ delays a bit further the transition to the CDM behavior, and the changes in the acoustic peaks are even more pronounced, as it can be seen in Figure~\ref{tps} (right) and Figure~\ref{highl}.

Thus, considering numerical solutions with a difference within sub-percent levels with respect to CDM, and with lower amplitude that the minimum sensitivity of CMB experiments, the range of axion masses with $\lambda\neq 0$ consistent with CMB observations seem to be given by $m_\phi > 10^{-24}$eV. This will be important in Section \ref{mp} when we carry out the statistical analysis.

\subsection{Mass power spectrum \label{sec:mps}}
\begin{figure*}[h!]
            \centering 
            \includegraphics[width=0.8\textwidth]{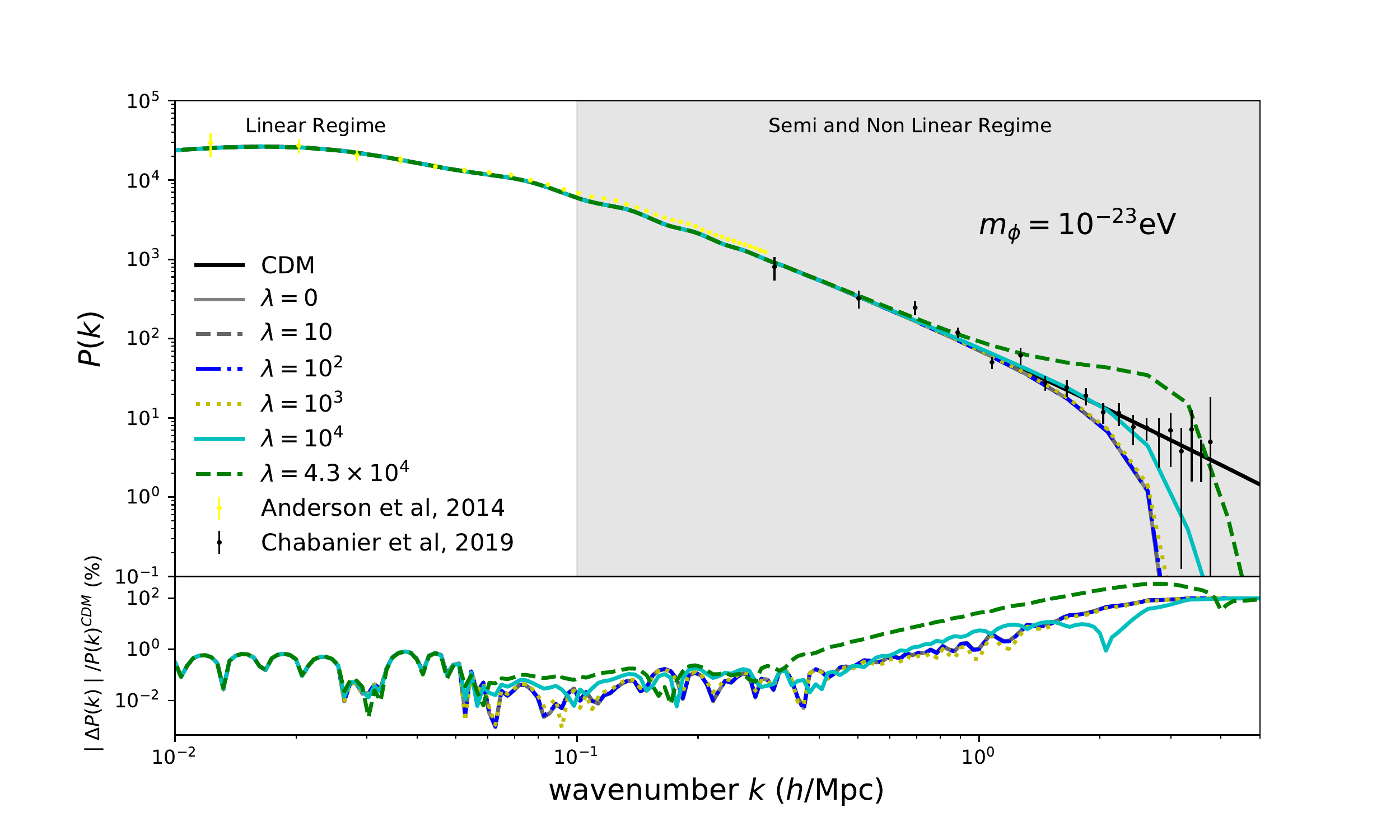}
            \includegraphics[width=0.8\textwidth]{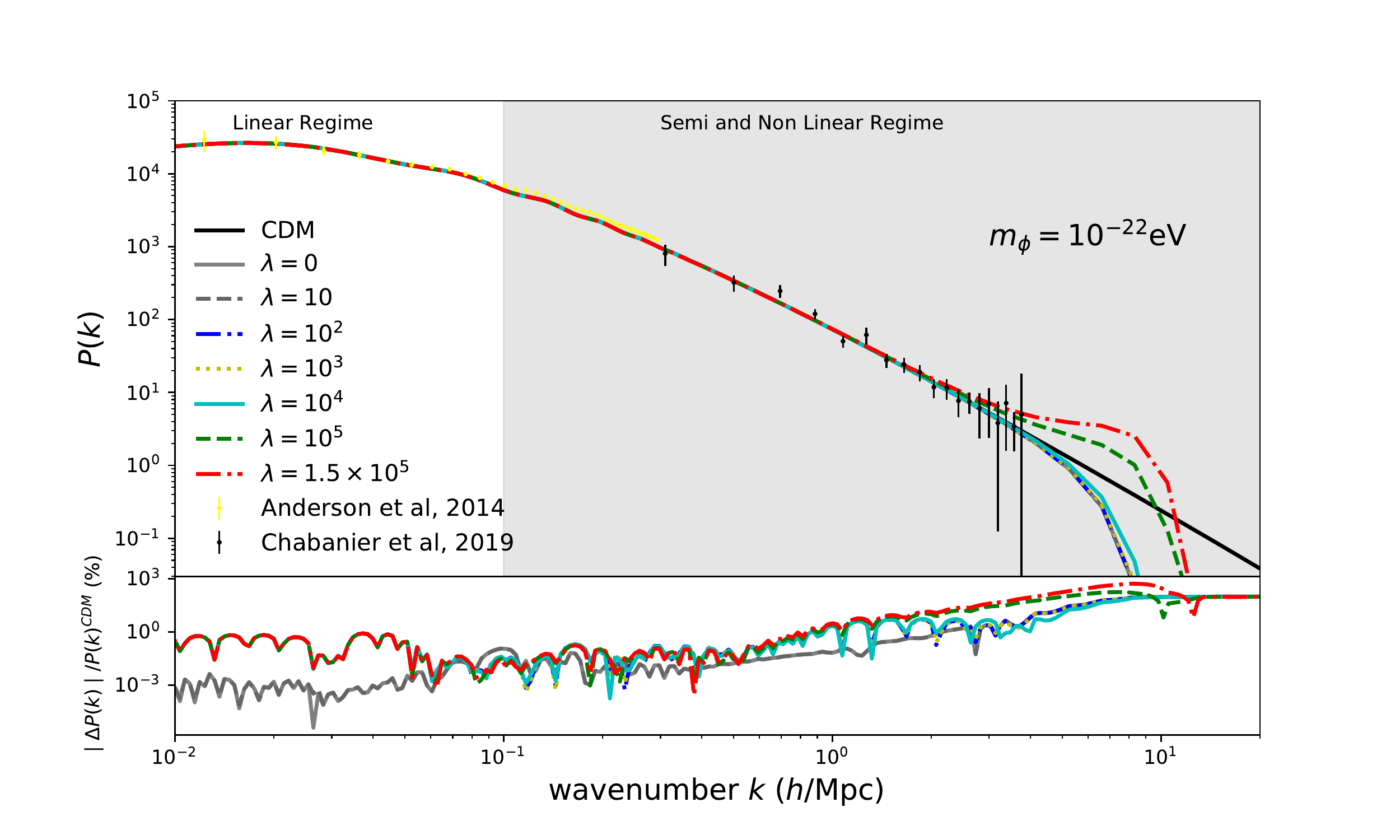}
        \caption[ mps ]
        {MPS for SFDM with axion masses $m_\phi/$eV$=10^{-22},10^{-23}$, and $\lambda$ from zero up to the maximum values reached for each axion mass. It can be noted that, for all the axion masses considered there is a cut-off at small scales (larger $k$'s), and even more, there is an enhancement of the MPS at such scales when considering large values of the parameter $\lambda$. Cosmological data from BOSS DR11 (yellow dots) \cite{Anderson:2013zyy}, and from Ly$\alpha$ forest (black dots) \cite{Chabanier:2019eai} are shown for reference.} 
        \label{mps}
    \end{figure*}

Figure~\ref{mps} shows the MPS for CDM (black line) and SFDM with masses $m_\phi =10^{-22},10^{-23} \mathrm{eV}$ for several values of the decay parameter $\lambda = 0, 10^1, 10^2, 10^3, 10^4$ (solid gray, dashed gray, dashed blue, dotted yellow, and solid cyan curves respectively), as well as for the extreme values of $\lambda$ corresponding to each value of the axion mass $\lambda = 4.3\times 10 ^4$ for $m_\phi =10^{-23}$eV (dashed green line), and $\lambda = 10^5, 1.5\times 10^5$ for $m_\phi =10^{-22}$eV (dashed green, and dotdashed red curves, respectively). For each case we observe the well-known cut-off at large wavenumbers, but this time there is also present a bump in the MPS at the cut-off scale for each value of the axion mass. As discussed in Sec.~\ref{blp}, the tachyonic instability produces an enhancement in the density contrast after the onset of the oscillations of the axion field, and such instability is going to be present in the MPS, at least for a range of wavenumbers as explained in~\citep{Cedeno:2017sou} and in Section~\ref{ti}. It is important to note that, for the case $m_\phi=10^{-22}$eV with $\lambda=10^5$ (green dashed curve in the bottom plot) the bump is within the range of wavenumbers shown in Figure~\ref{dcomega2}. For a qualitative comparison, we have included data from BOSS DR11 (yellow dots) \cite{Anderson:2013zyy}, and from Ly$\alpha$ forest (black dots) \cite{Chabanier:2019eai}, which show that these data, specially the latter (see below), can in principle constrain the presence of both, the bump and cut-off in the MPS.

\subsection{Lyman-alpha}\label{ly_const}
Based on the comparison of the CMB anisotropies and MPS with available data, we see that the SFDM model describes such cosmological observables as good as CDM model does, as long as the axion mass is $m_\phi \gtrsim 10^{-24}$eV. This is a lower value than that imposed by Lyman-$\alpha$ observations of the 1-dimensional flux power spectrum ($P^{1D}$), for the axion mass endowed with a quadratic potential (FDM case), given by $m_\phi \gtrsim 10^{-21}$eV \cite{Irsic:2017yje,Armengaud:2017nkf}\footnote{Note that \cite{Zhang:2017chj} reports a different constraint with the same observations, claiming that including quantum pressure to numerical simulations of FDM leads to a lower bound of $m_\phi = 10^{-23}$eV.}.

\begin{center}
\begin{figure}[h!]
    \centering
   \includegraphics[width=0.49\linewidth]{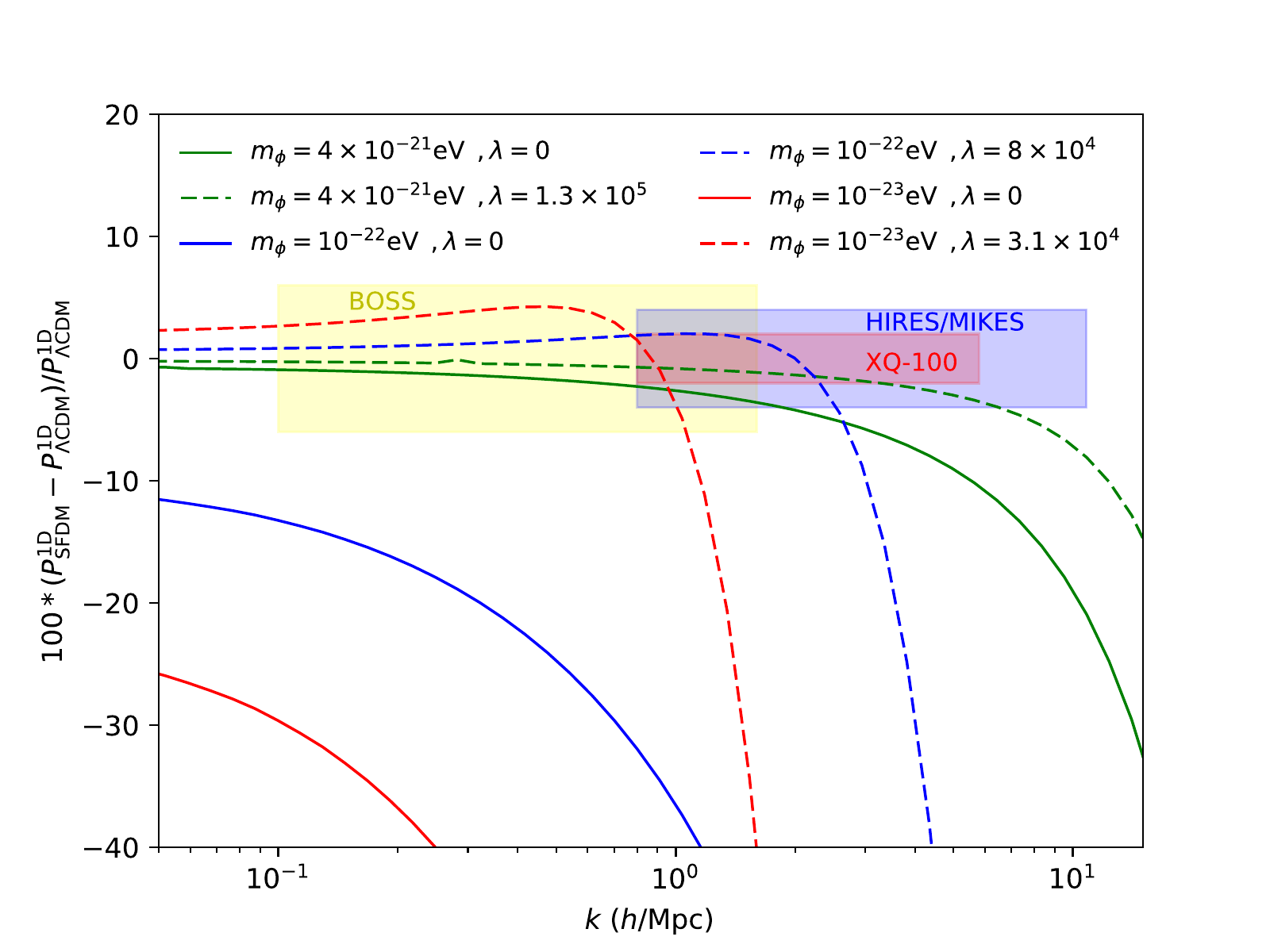}
        \label{fig:first_sub}   
    \caption{1D MPS for the axion field compared to the $\Lambda$CDM one. We show the cases $m_{\phi}=4\times 10^{-21} {\rm eV}$ for $\lambda=0,1.3 \times 10^5$ (green lines), $m_{\phi}=10^{-22} {\rm eV}$ for $\lambda=0, 8 \times 10^4$ (blue lines), and $m_{\phi}=10^{-23} {\rm eV}$ for $\lambda=0, 3.1 \times 10^4$ (red lines). For reference we have included  colored rectangles indicating the rough precision of current data from BOSS \cite{Palanque-Delabrouille:2013gaa} (yellow), HIRES/MIKES \cite{Viel:2013apy} (blue) and XQ-100 \cite{2016A&A...594A..91L} (red) to show that these experiments can be used to constraint the axion field parameters $m_\phi$ and $\lambda$.}
    \label{ps1d}
\end{figure}
\end{center}

To qualitatively assess the constraints that Lyman-$\alpha$ can impose upon SFDM, we compare in Figure~\ref{ps1d} the relative difference with respect to CDM, for the  $P^{1D}$ with the precision of current measurements from data sets such as eBOSS \cite{Palanque-Delabrouille:2013gaa}, HIRES/MIKES\cite{Viel:2013apy} and XQ-100 \cite{2016A&A...594A..91L} (yellow, blue and red rectangle respectively). We do this for the following combinations: $m_{\phi}=4\times 10^{-21} {\rm eV}$ for $\lambda=0,1.3 \times 10^5$ (green curves), $m_{\phi}=10^{-22} {\rm eV}$ for $\lambda=0, 8 \times 10^4$ (blue curves), and $m_{\phi}=10^{-23} {\rm eV}$ for $\lambda=0, 3.1 \times 10^4$ (red curves). We can see that combinations with $\lambda=0$ are excluded by the data except for the larger masses, while combinations with $\lambda > 0$ seem to be preferred. This means that an axion field endowed with a trigonometric potential could still be allowed by Lyman-$\alpha$ observations (like for instance the future data from DESI~\cite{levi2013desi}).

\subsection{Comparison with Data\label{mp}}
We will use the parameter estimator code \textsc{Monte Python}~\cite{audren2013conservative} to analyze the parameter space of our model and imposing constraints by using CMB observations from the latest data release of the Planck Collaboration 2018~\cite{Aghanim:2018eyx,Aghanim:2019ame}. We have also built up a likelihood from the Lyman-$\alpha$ forest data reported by~\cite{Chabanier:2019eai}, in order to impose constraints on our model from small scale structures. It is important to mention that what we are using is a Lyman-$\alpha$ 3D MPS at redshift $z=0$, which the authors in~\cite{Chabanier:2019eai} have inferred from the 1D flux power spectrum measured by the BOSS and eBOSS collaboration~\cite{Chabanier:2018rga}. Inferring the linear matter spectrum at $z=0$ is a highly model--dependent process, and therefore, the constraints obtained in this work from Lyman-$\alpha$ have to be understood as how much SFDM can deviate from the CDM case, which is the fiducial model considered in~\cite{Chabanier:2019eai}, and should only be read as an approximation for what is the effect of having the full axion potential on the constraints for the axion mass.

Our model is defined by two parameters, the axion mass $m_\phi$ and the decay parameter $\lambda$, and additionally by the standard cosmological parameters of $\Lambda$CDM model: the physical baryon density parameter $100\omega_b$, the (logarithmic) power spectrum scalar amplitude $\log(10^{10}A_s)$, the scalar spectral index $n_s$, the Thomson scattering optical depth due to reionization $\tau_{reio}$, and the angular size of sound horizon at decoupling $100\theta_s$. Note that we do not include $\Omega_c$ (dark matter density parameter) because that information will be provided by our axion field. Thus, in principle we have in total 7 cosmological parameters given by $\Theta = \left[ 100\omega_b, \log(10^{10}A_s), n_s, \tau_{reio}, 100\theta_s, \log m_\phi, \log \lambda \right]$, where we have defined the scalar field parameters $m_\phi$ and $\lambda$ in logarithmic scale. However, from previous analysis we obtained that the posteriors of the standard parameters remain basically unchanged with respect to the $\Lambda$CDM case, and then we will focus on presenting the results of the two parameters of our model, i.e., $\Theta = \left[\log m_\phi, \log \lambda \right]$.

The initial input given to the code to run the chains is summarized in Table \ref{input}, where the initial mean value, as well as the priors and the 1-$\sigma$ value, are specified for each of the parameters $\Theta$. The input for the axion field parameters $m_\phi$ and $\lambda$ are chosen to be consistent with the numerical solutions obtained with \textsc{class}. Thus, the means and priors for $m_\phi$ and $\lambda$ will be set based on the cosmological evolution of the axion field that we were able to explore numerically.
\begin{table}[h!]
\centering
\begin{tabular}{|l|c|c|c|c|}
 \hline
Param & mean & prior min & prior max & 1-$\sigma$ \\ \hline
$\log \lambda$ & 5 & 0 & 10 & 0.05 \\
$\log m_\phi$ & -21 & -26 & -16 & 0.05 \\
\hline
 \end{tabular}
\caption{Initial input for the parameters $\Theta$ of our SFDM model. The prior for the axion field parameters were chosen according to the numerical solution we obtained from \textsc{class}.}
\label{input}
\end{table}

\begin{center}
\begin{figure}[h!]
    \centering
   \includegraphics[width=0.7\linewidth]{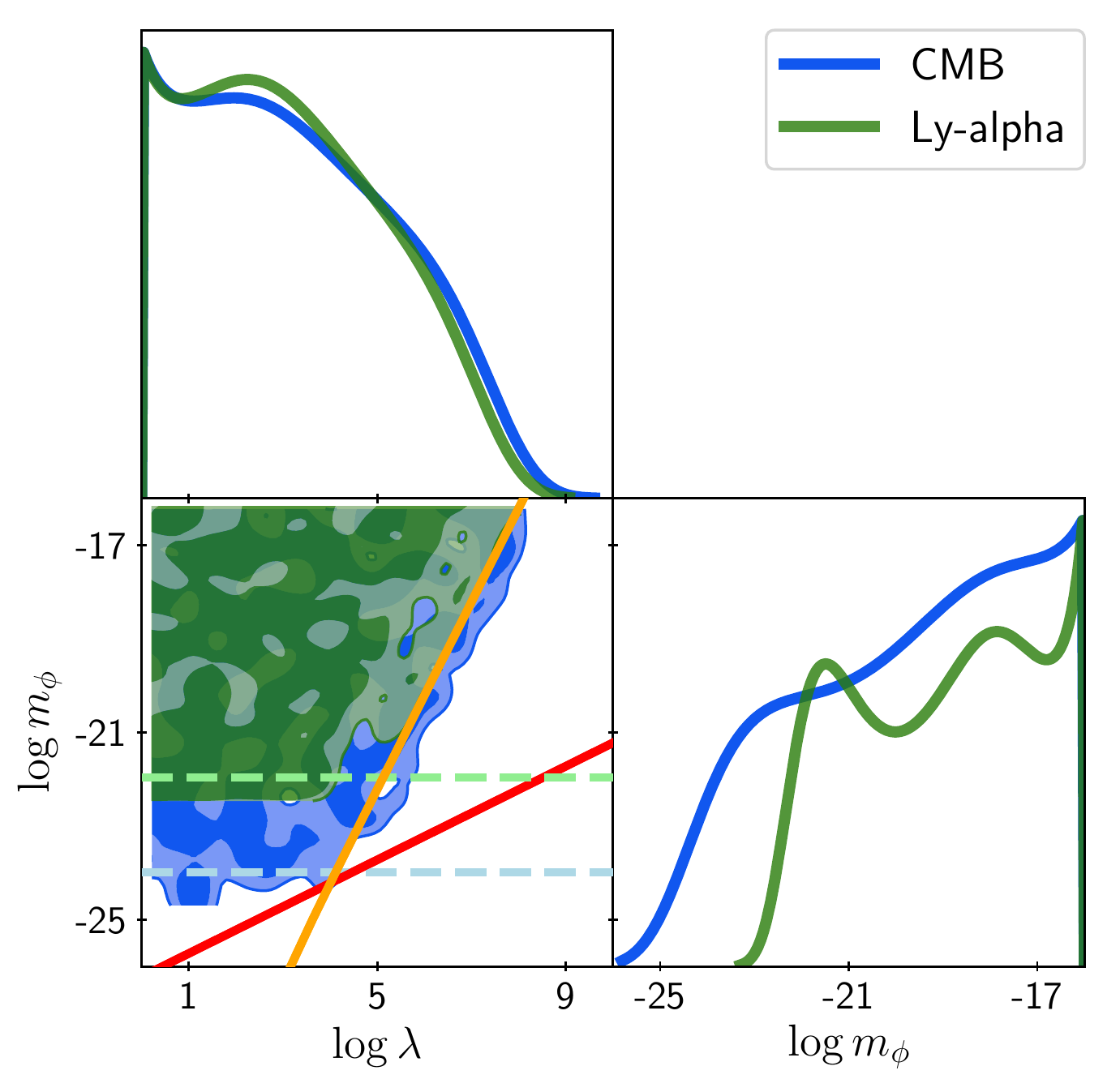}
    \caption{1D and 2D posterior distributions for the axion field parameters $m_\phi$ and $\lambda$ in logarithmic scale. When considering the CMB observations from Planck 2018 (blue), we can set a lower bound for the value of the axion mass of $\log m_\phi = -23.99$ at $95.5\%$ C.L. (blue dashed line). On the other hand, Lyman-$\alpha$ data (green) imposes a stronger constraint on the mass given by $\log m_\phi = -21.96$ at $95.5\%$ C.L. (green dashed line). Orange and red lines in the 2D posterior represent numerical and physical limits respectively. See text for more details.}
    \label{cp}
\end{figure}
\end{center}

We have run the chains with the Metropolis-Hasting algorithm, with the Gelman-Rubin convergence criterion~\cite{Gelman:1992zz} fulfilling $R-1<0.05$. The minimum of the likelihood and the $\chi^2$ function we obtained are respectively given by $-\ln{\cal L}_\mathrm{min} =395.423$, $\chi^2_{\rm{min}}=790.8$ from CMB, and $-\ln{\cal L}_\mathrm{min} =10.7182$, $\chi^2_{\rm{min}}=21.44$ from Lyman-$\alpha$. The posteriors are shown in Figure \ref{cp}, where it can be seen that the axion field parameters $m_\phi$ and $\lambda$ have a non--Gaussian posterior. However, we observe that the axion mass shows a lower bound given by $\log m_\phi = -23.99$ at $95.5\%$ C.L when considering CMB observations (blue dashed line). This is consistent with the previous result shown in Section~\ref{cmb_anisot}, where we compare our numerical solutions with data from the CMB anisotropies (see Figure~\ref{highl}).

Data from Lyman-$\alpha$ forest impose a stronger constraint on the axion mass, given by $\log m_\phi = -21.96$ at $95.5\%$ C.L. (green dashed line). This is expected since, as we mentioned before, it is well--known that Lyman-$\alpha$ observations are more restrictive on the value of the axion mass in comparison with the bounds obtained from CMB. A recent constraint of a 95\% lower limit $m_{\phi}>2\times 10^{-20}$eV has been reported in~\cite{Rogers:2020ltq}. In such work the authors use Lyman-$\alpha$ forest to find bounds on the axion mass by studying the suppression of cosmic substructures, which is modeled by an analytical expression of the transfer function $T(k)$ previously proposed in~\cite{Murgia:2017lwo,Murgia:2017cvj,Murgia:2018now}. In appendix~\ref{abg} we explain in detail why this type of analysis, as proposed in the references above, can not be done directly for the model in turn and therefore we opted to work with the 3D MPS from \cite{Chabanier:2019eai} as a proxy to find how constraints to the axion mass relaxes when we consider the full axion potential. More reliable constraints to the axion mass should arise from a dedicated analysis of the Lyman alpha forest observations.

Therefore, the presence of $\lambda$ in our analysis plays a key role in the axion mass constraint: it allows to the SFDM model with a fiducial mass of $m_{\phi}\simeq 10^{-22}$eV to be in agreement with the bounds impose by Lyman-$\alpha$ observations, as we have anticipated from the $P^{1D}$ in Section~\ref{ly_const} (see Figure~\ref{ps1d}). In this sense, whereas CMB observations impose a direct bound on $m_{\phi}$, in the case of Lyman-$\alpha$ forest what we have is the most likely values for the axion mass such that they are in agreement with the $\Lambda$CDM--based data from the inferred Lyman-$\alpha$ 3D MPS.

We also show in Figure~\ref{cp} two limits: (1) a numerical one given by the extreme case of $\lambda$ for a given axion mass $m_{\phi}$ (orange line), and (2) a physical one given by the condition for the SFDM to provide the total DM (red line). Limit (1) allows us to see that what seems to be a constraint on the decay parameter $\lambda$ is, in fact, a manifestation of the numerical limitations of \texttt{CLASS} to handle large values of $\lambda$ of a given mass (see table~\ref{table} of Appendix~\ref{sec:ewdm}). Thus, the region at the left of the orange line contains all possible combinations of $\left\lbrace m_{\phi}\, , \lambda \right\rbrace$ that are numerically allowed. Limit (2) is given by $m_{\phi}/\sqrt{\lambda}>6\times 10^{-27}$eV~\cite{Diez-Tejedor:2017ivd,Cedeno:2017sou}, which in the (logarithmic) $\left\lbrace m_{\phi}\, , \lambda\right\rbrace$--plane is represented by the upper region to the straight red line $\log m_{\phi} = (1/2)\log\lambda - 26.22$. Thus, the constraints on the axion mass that we have found are consistent with a SFDM model where all the DM budget is provided by the axion field.

\section{Halo formation within axion models \label{sec:halos}}
To explore other possible deviations from the CDM model on such cosmological quantities, in this section we present an approach to obtain the evolution of both, the growth factor $D$ and the velocity growth factor $f$ as function of the wavenumber $k$ for SFDM with the axion--like potential. We will also study the \textit{Halo Mass Function} (HMF), which encodes the comoving number density of DM halos as function of the halo mass, and it constitutes a representative cosmological probe of DM and dark energy. It can be used for example to constraint the value of the combined parameters $\sigma_8$ and $\Omega_M$ (the power spectrum normalization and the matter density parameter, respectively), and also to characterize the dark energy equation of state~\cite{Vikhlinin:2008ym,doi:10.1146/annurev-astro-081710-102514,Murray:2013sna}.

\subsection{Growth factor $D$ and velocity growth factor $f$ with scale-dependence \label{dkfk}}
It is well known that the growth factor $D$ for CDM model is independent of the wavenumber $k$, but it is the transfer function $T(k)$ which carries such information. For instance, one can write for the gravitational potential sourced by density perturbations $\Phi(k,a) \propto T(k)D(a)$, which simplifies the study for the growth of matter overdensities in the CDM scenario~\cite{heath1977growth,Lahav:1991wc,Eisenstein:1997ij,Dodelson:2003ft,Huterer:2013xky}. Likewise, the standard parameterization for the velocity growth factor does not contain explicit information of $k$, $f(z)=\Omega_m^{\gamma}(z)$~\cite{wang1998cluster,Linder:2007hg,Polarski:2007rr,Guzzo:2008ac,Gannouji:2008jr,Basilakos:2019hlb,Khyllep:2019odd}, where $\gamma$ is called the growth index, and $\Omega_m$ is the energy density parameter for the total matter as function of the redshift $z$. However, the scale-dependence on the quantities $D$ and $f$ have been studied in alternatives models of gravity~\cite{Tsujikawa:2007gd,Gannouji:2008wt,Fu:2010zza,Raccanelli:2012gt,Huterer:2013xky,LopezRevelles:2013vm,Nesseris:2015fqa,Nesseris:2017vor} mainly due to the appearance of an effective Newton's constant containing explicit dependence on $k$. In the axion case, the scale-dependence is present already in the equations of motion of density perturbations, and then one requires to study their numerical solutions for a better understanding of their growth.

As starting point, let us revisit the system of equations that rules the dynamics of the SFDM linear perturbations after the onset of rapid oscillations. From Eq.~(\ref{eqnewdeltas}), and considering $\cos \theta \sim \sin \theta \sim 0$, we find
\begin{equation}
\delta_0'' + \omega^2 \delta_0 = -\frac{\bar{h}^{\prime \prime}}{2} + 2 \frac{k^2}{k_J^2}\frac{k_J^{\prime}}{k_J} \delta_1 \, ,
\label{ded0}
\end{equation}
where, in contrast to Eq.~\eqref{eq:harmonic}, we are not neglecting the evolution of the Jeans wavenumber $k_J$. Two main features can be seen in Eq.~\eqref{ded0}: 1) the solution of $\delta_0$ will always be coupled to $\delta_1$, and 2) the solution for $\delta_0$ will depend on the wavenumber $k$.

\begin{center}
\begin{figure}[htp!]
    \centering
   \includegraphics[width=0.49\linewidth]{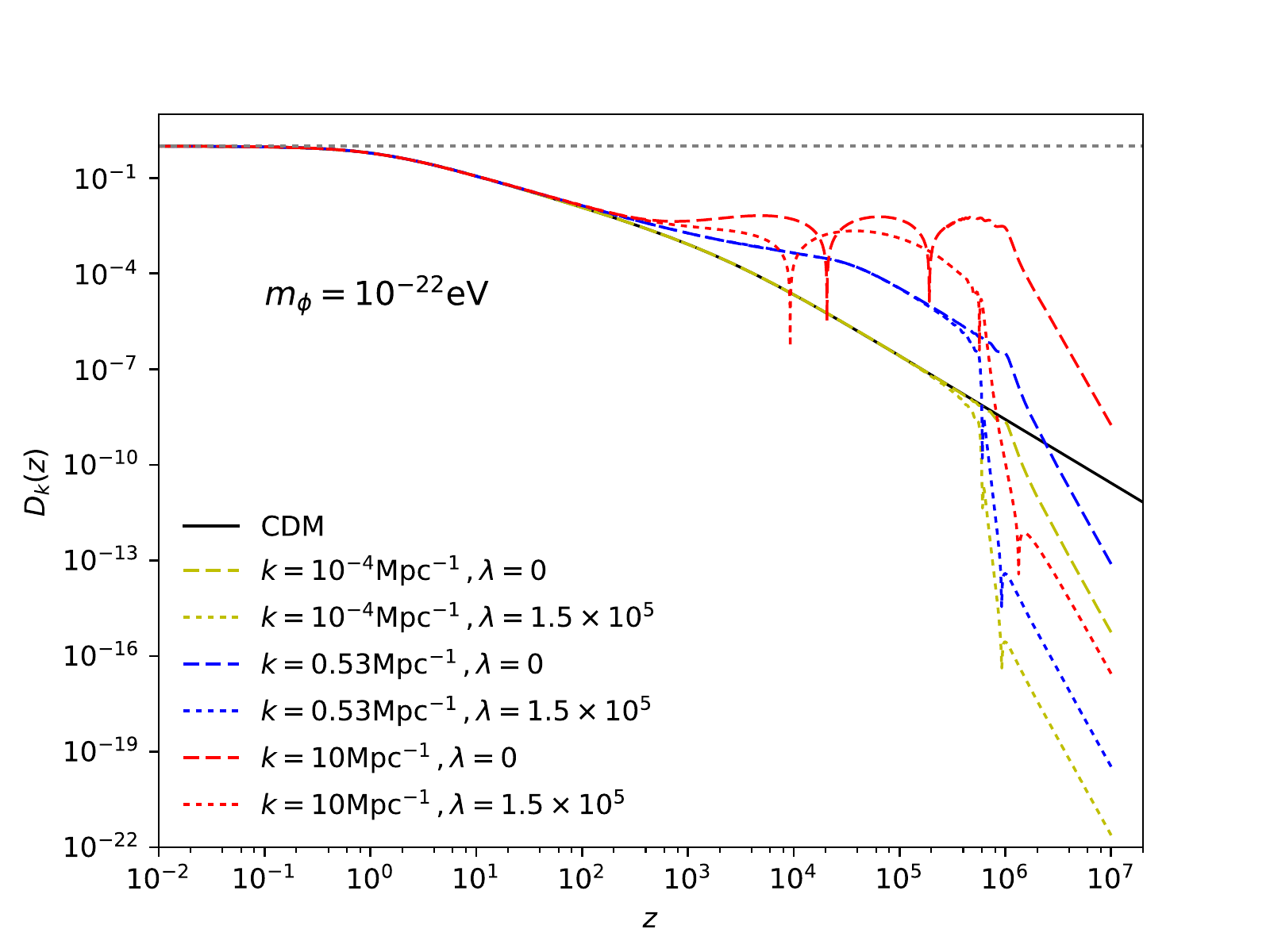}
        \label{fig:first_sub}   
    \caption{Growth factor $D_k(z)$ for an axion mass $m_\phi = 10^{-22}$eV with both, quadratic potential (dashed lines) and trigonometric potential (dotted lines). The tachyonic instability is manifested for the latter as a fast increment of amplitude for $D_k$ at $z\sim 10^6$. Horizontal dotted gray line indicates $D=1$, where all curve converge at $z\sim 0$.}
    \label{gfplot}
\end{figure}
\end{center}

Following recent literature, where the growth factor is defined in terms of the density contrast \cite{Sapone2007:0709.2792v1,Gong2009:0903.0001v2,Lee2009:0907.2108v2,Acquaviva:2010vr,Zheng2010:1010.3512v3,Porto2011:1101.2453v1,Huterer:2013xky,Marsh:2013ywa,Zhai2017:1705.10031v2,Basilakos:2019hlb}, we define a scale-dependent growth factor $D_k$ as
\begin{equation}
D_k(z) \equiv \frac{\delta_0(z,k)}{\delta_0(z=0,k)}\, ,
\label{gf}
\end{equation}
so that $D_k(z=0) = 1$. The definition given in Eq.~\eqref{gf} allows us to generalize the growth factor in such a way that it is possible to track its evolution for each wavenumber $k$. This is done in Figure~\ref{gfplot}, where we show the growth factor $D_k(z)$ for $k=10^{-4}$Mpc$^{-1}$ (yellow), $k=0.53$Mpc$^{-1}$ (blue), $k=10$Mpc$^{-1}$ (red). The axion mass is $m_{\phi}=10^{-22}$eV, and we also show the cases of the quadratic potential (FDM case $\lambda=0$, dashed lines) and the trigonometric one ($\lambda = 1.5\times 10^5$, dotted lines). The initial amplitude for the growth factor with trigonometric potential is smaller than that of the FDM case, but around $z \sim 10^6$ the growth factor with $\lambda = 1.5\times 10^5$ suffers the tachyonic instability and its amplitude start to grow faster. It is important to recall that such fast growth is translated as a bump in the linear MPS, as was shown in Figure \ref{mps}. Interestingly enough, from $z\sim 100$ up to the present day, all curves evolve as CDM, which implies that for $z<100$ the growth factor $D_k(z)$ in Eq.~\eqref{gf} becomes effectively scale-independent.

\begin{center}
\begin{figure}[h!]
    \centering
   \includegraphics[width=0.49\linewidth]{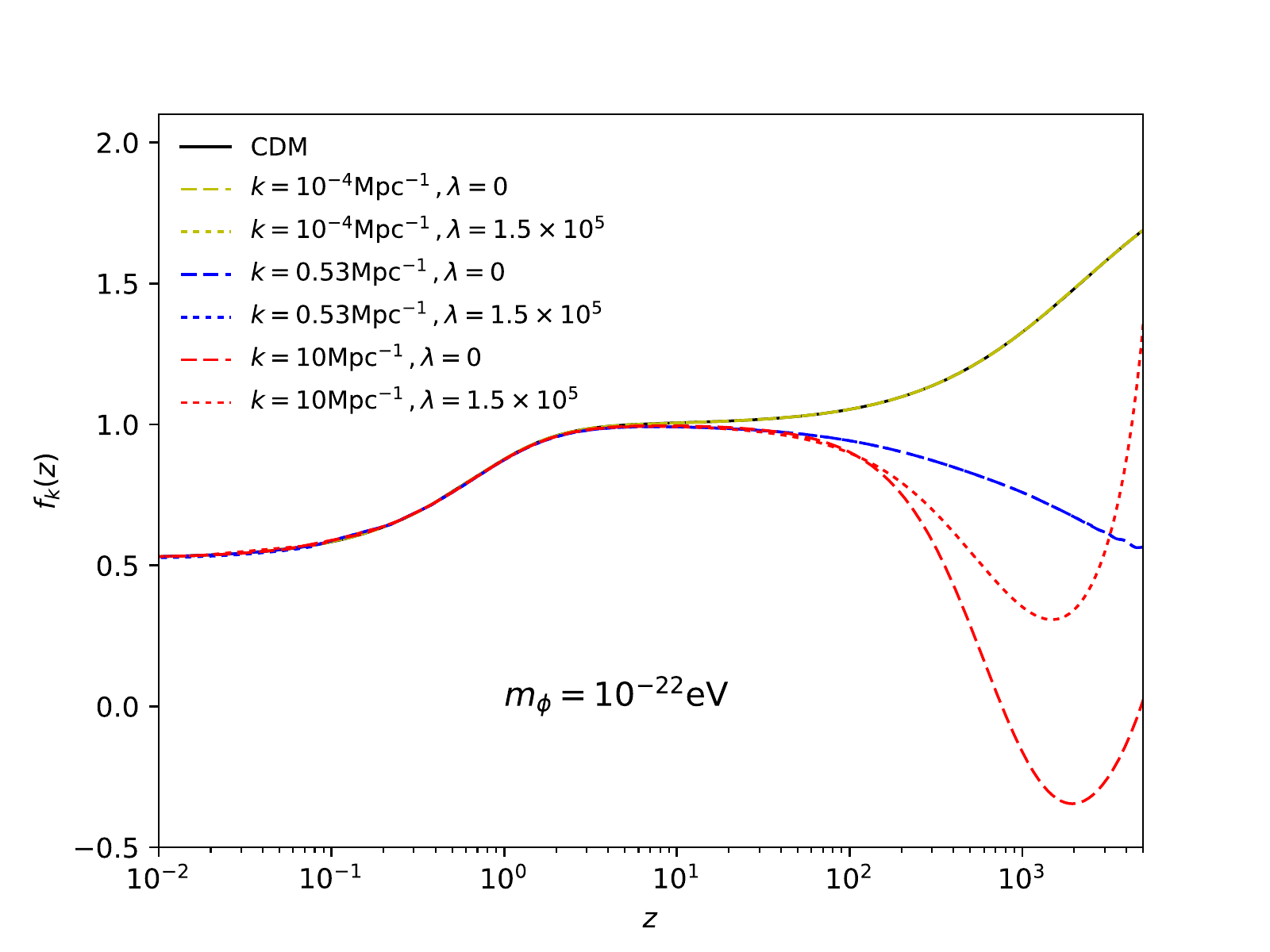}
        \label{fig:first_sub}   
    \caption{Velocity growth factor $f_k(z)$ for an axion with mass $m_\phi=10^{-22}$eV. Dashed (dotted) lines correspond to $\lambda=0\, (\lambda=1.5\times 10^5)$, and yellow, blue and red lines indicate wavenumbers $k=10^{-4}, 0.53, 10$Mpc$^{-1}$ respectively. For $k\ll 1$Mpc$^{-1}$ the velocity growth factor evolves as CDM for all redshift, whereas for $k>1$Mpc$^{-1}$ each mode evolve independently until $z\sim 10$, where all curve converge to the CDM case, and the velocity growth factor is the same for all wavenumbers. See text for more details.}
    \label{vgfplot}
\end{figure}
\end{center}

Going further, the definition given by Eq.~\eqref{gf} enables us to write the velocity of the growth factor $f_k(z)$ as follows,
\begin{equation}
f_k(z) = \frac{d\log D_k(N)}{dN} = -(1+z) \frac{d\log D_k(z)}{dz} = - (1+z) \frac{d\log \delta_0(z,k)}{dz} \, .
\label{fk}
\end{equation}
The dependence on $k$ for the function shown above can be seen in Figure~\ref{vgfplot}, where the colors and the line style for each curve are the same as in Figure~\ref{gfplot}. Notice that the velocity growth factor for $k=10^{-4}$Mpc$^{-1}$ is the same as that of CDM and is not affected by the values of $\lambda$; that is, at large scales we recover the same behavior of CDM. Similarly, for $k=0.53 \, \mathrm{Mpc}^{-1}$ the evolution is also independent of the values of $\lambda$, although the CDM evolution is not recovered for $z\gtrsim 10$. Thus, it is possible to distinguish between CDM and SFDM at high redshifts. The result is different for the wavenumber $k=10$Mpc$^{-1}$, where we can see that the evolution of $f_k$ is sensitive to the two values of $\lambda$ considered. However, from $z\sim 10$ to the present day, the evolution of $f_k$ for each mode and for each value of $\lambda$ is the same as that of CDM. This means that at late times the MPS of the axion field should keep a constant ratio with respect to that of CDM. 

\begin{center}
\begin{figure}[h!]
    \centering
   \includegraphics[width=0.49\linewidth]{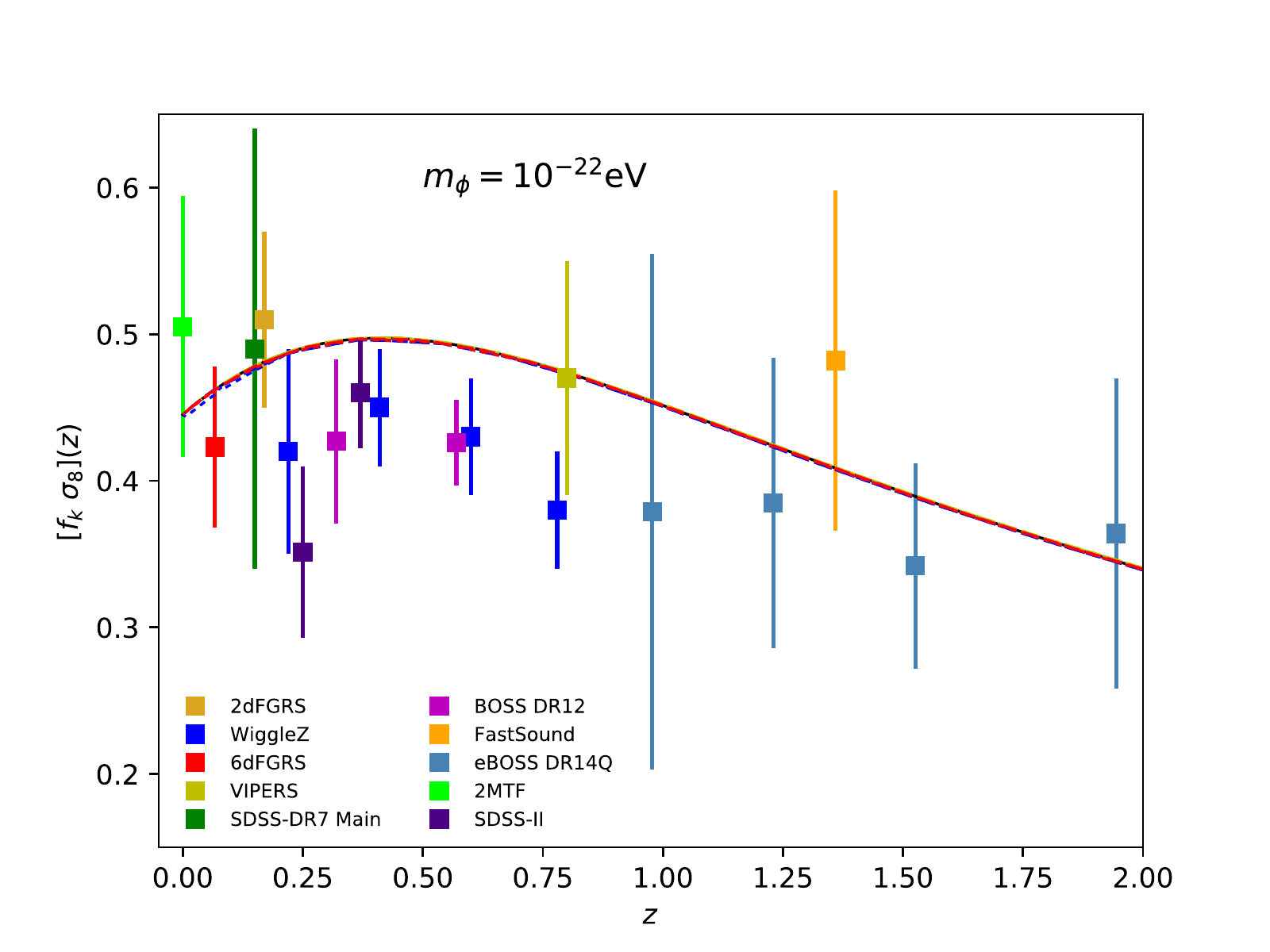}
        \label{fig:first_sub}   
    \caption{Velocity growth factor $f_k$ and variance $\sigma_8$ combined as function of both, wavenumber $k$ and redshift $z$. The overlapped curves have the same values of $k$ and $\lambda$ as those of the previous Figures \ref{gfplot} and \ref{vgfplot}. Observational data are shown in colored squares from 2dFGRS \cite{Song:2008qt}, WiggleZ \cite{Blake:2011rj}, 6dFGRS \cite{Beutler:2012px}, VIPERS \cite{delaTorre:2013rpa}, SDSS DR7 Main \cite{Howlett:2014opa}, BOSS DR12 \cite{Gil-Marin:2016wya}, FastSound \cite{Okumura:2015lvp}, eBOSS DR14Q \cite{Zhao:2018gvb}, 2MTF \cite{Howlett:2017asq} and SDSS-II \cite{Samushia:2011cs}.}
    \label{fksk}
\end{figure}
\end{center}

For $k=10$Mpc$^{-1}$, we attribute the notorious difference at $z>10$ between the FDM case and the axion-like potential to the tachyonic instability, since this effect is manifested at such range of scale. Finally, since the growth factor $D_k$ and the velocity growth factor $f_k$ coincide with those of CDM for $0<z<10$, the combined observable $f_k\sigma_8$ at $0<z<2$ (range within which we can search for observational constraints) will be insensitive to the details of the axion case, as can be seen in Figure~\ref{fksk}, where the overlapped curves correspond to the same values of wavenumbers $k$ and decay constant $\lambda$ as those in Figures~\ref{gfplot} and~\ref{vgfplot}.

Whereas strong constraints have been imposed to the SFDM  mass $m_{\phi}$ (through galactic observations, MPS and CMB anisotropies), having observations of matter distribution at high redshifts can be useful to explore the nature of DM, and particularly to constraint the decay parameter $\lambda$ of the axion field. The cosmological effects of such parameter have not be studied in great detail, and we are showing that it has a characteristic imprint on the structure formation, at small scales (see MPS in Figure~\ref{mps}) as well as at high redshifts (figures~\ref{gfplot} and~\ref{vgfplot}).

\subsection{Semi-analytical Halo Mass Function\label{mf}}
A halo is an overdensity of matter, which lie on the non--linear regime of structure formation. To study such objects numerical simulations have to be carried out \cite{Jenkins:2000bv,Tinker:2008ff,Warren:2005ey}. However, semi-analytical analysis can be performed as well, as have been shown in \cite{Press:1973iz,Bond:1990iw,Sheth:1999mn,Sheth:1999su}. Particularly, the procedure to obtain the semi analytical HMF of our model will be similar to that given by~\cite{Schneider:2013ria,Du:2016zcv}. 

First, we define the window functions we are going to implement: the Top-Hat window function $W_{TH}$, which is a filter with spherical symmetry in real space, and the Sharp-k window function $W_{SK}$, defined as a Top-Hat function in Fourier space. They are given, in Fourier space, by
\begin{equation}
W_{TH}(kr) = \frac{3}{(kr)^3}\left[ \sin(kr) - kr\cos(kr) \right]\, ,\quad W_{SK}(kr) = \Theta(2\pi - kr)\, .
\label{wf}
\end{equation}
The Top-Hat function is useful to work with the CDM model, while the Sharp-k function is useful for suppressed power spectra, which is the case of the axion field. Whereas the mass for the Top-Hat is defined as usual $M_{TP}=4\pi {\bar{\rho}} R^3/3$, in the case of the Sharp-k we have that $M_{SK}=4\pi {\bar{\rho}} (cR)^3/3$, where $c$ is a free parameter usually fixed by numerical simulations. In both cases, ${\bar{\rho}}$ is the average density of the universe~\cite{Press:1973iz}. More discussion about the choice of the window functions are given in \cite{Urena-Lopez:2015gur,Benson:2012su,Schneider:2014rda,Buckley:2014hja} and references therein.
\begin{center}
\begin{figure}[htp!]
    \centering
   \includegraphics[width=0.49\linewidth]{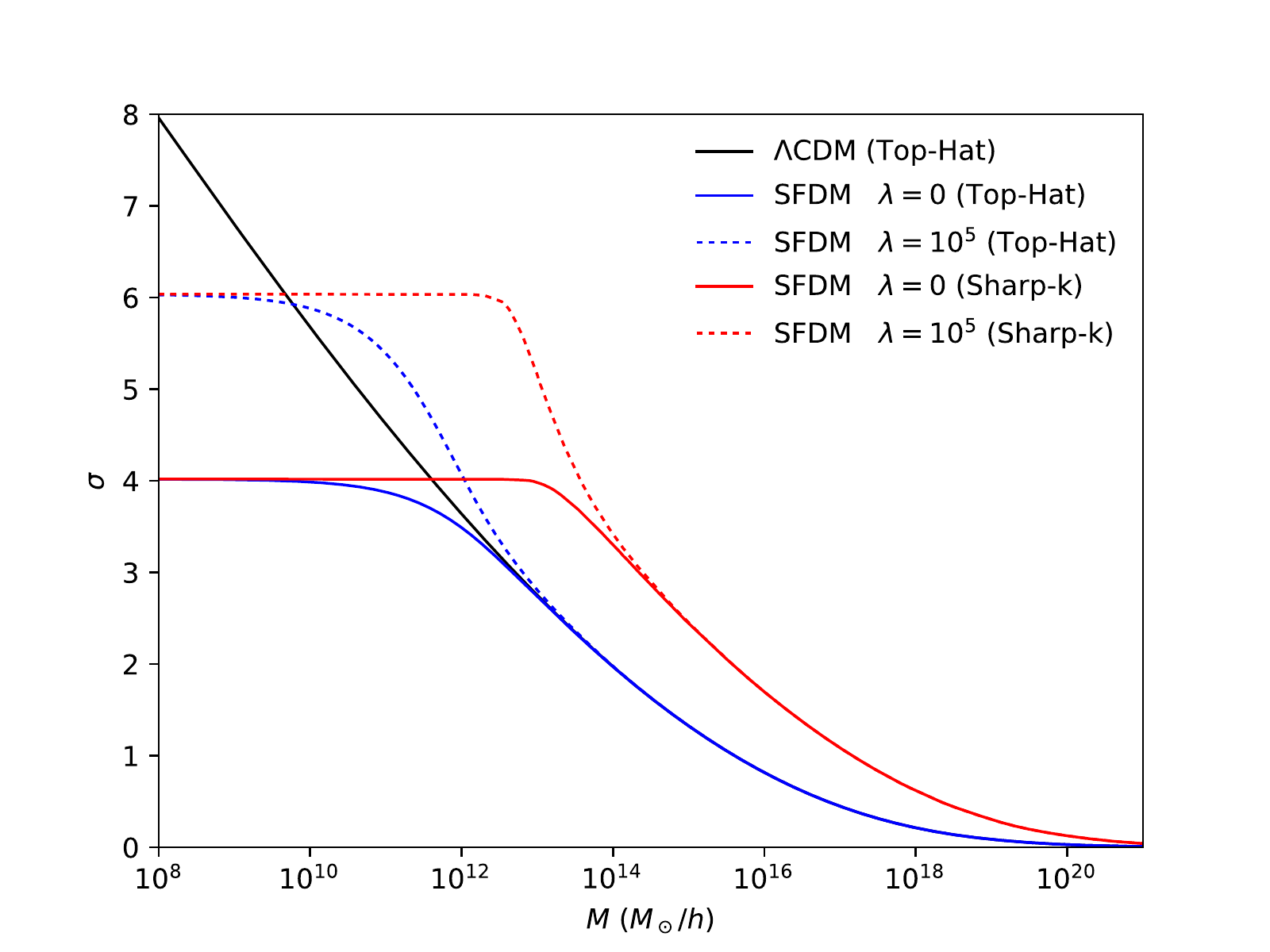}
        \label{fig:first_sub}   
    \caption{Square root of the variance at $z=0$ as function of the halo mass for $\Lambda$CDM (solid black line) and SFDM for $\lambda=0$ (solid lines) and $\lambda=10^5$ (dotted lines) with Top-Hat (blue lines) and Sharp-k (red lines) window functions respectively.}
    \label{sigma}
\end{figure}
\end{center}

One of the quantities of interest is the \textit{variance}, which is calculated as
\begin{equation}
\sigma^2(r) = \int \frac{d^3\vec{k}}{(2\pi)^3}P(k)W^2(kr)\, .
\end{equation}
In Figure \ref{sigma} we show the square root of the variance at redshift $z=0$ for CDM and SFDM, the latter with quadratic ($\lambda = 0$) and trigonometric ($\lambda = 10^5$) potentials, and $m_\phi = 10^{-22}$eV. The variance of the axion field, for both the Top Hat and Sharp-k window functions, show a constant value for small halo masses, in contrast to that of the CDM model, which is always increasing. The asymptotic values for the quadratic and trigonometric potentials are different; for the latter it can be seen that it is the tachyonic instability, and ultimately the non-linearities of the trigonometric potential, that enhances the value of $\sigma$ at small halo masses.

To study the gravitational collapse within SFDM, it is necessary to take into account the scale-dependence of its density perturbations. The threshold value at which some matter fluctuation associated to a given mode $k$ will collapse, is known as the \textit{critical overdensity}, which within a standard CDM scenario is defined by \cite{Gunn:1972sv,Percival:2000aj,Percival:2005vm,Schneider:2013ria}
\begin{equation}
\delta_{crit} = 1.686\frac{D_{\rm{CDM}}(z=0)}{D_{\rm{CDM}}(z)}\, ,
\label{critovercdm}
\end{equation}
where $D_{\rm{CDM}}$ is the growth factor for CDM
\begin{equation}
D_{\rm{CDM}} = \frac{5\Omega_m H}{2}\int \frac{da}{a^3H^3}\, .
\end{equation}

In the case of SFDM, we can in principle apply a similar expression, but using the growth factor introduced in Eq.~\eqref{gf},
\begin{equation}
\delta_{crit} = 1.686\frac{D_k(z=0)}{D_k(z)}\, .
\label{critoversfdm}
\end{equation}
We are interested in building up the HMF at $z=0$, and even when Eq.~\eqref{critoversfdm} contains explicit dependence on the wavenumber $k$, the growth factor for SFDM coincides with the CDM case at later times, as was discussed in Section \ref{dkfk}. Therefore, this approach will not be useful to study the effects of gravitational collapsing with scale dependence on the HMF.

Notwithstanding, we can consider alternative approaches as those presented in~\cite{Marsh:2013ywa,Marsh:2016vgj,Du:2016zcv}, where the authors introduce a definition of the growth factor in terms of several density contrasts rates. Particularly, in Eq.~(10) from~\cite{Du:2016zcv}, the relative amount of growth between CDM and SFDM is written as
\begin{equation}
\frac{D_{\rm{CDM}}(z)}{D_{\rm{SFDM}}(M,z)} = \frac{\delta_{\rm{CDM}}(k,z)}{\delta_{\rm{SFDM}}(k,z)} \frac{\delta_{\rm{CDM}}(k_0,z_h)}{\delta_{\rm{SFDM}}(k_0,z_h)} \frac{\delta_{\rm{SFDM}}(k_0,z)}{\delta_{\rm{CDM}}(k_0,z)} \frac{\delta_{\rm{SFDM}}(k,z_h)}{\delta_{\rm{CDM}}(k,z_h)}\, ,
\label{critoverdu}
\end{equation}
where $k_0=0.002h/$Mpc is a pivot scale, and $z_h=300$ is the redshift at which the shape of the CDM power spectrum has frozen in. We observe that the pivot scale is small, and for such mode the growth factor of SFDM will evolve as CDM. Then, the second and third ratio in Eq.~\eqref{critoverdu} are $\delta_{{\rm{CDM}}}(k_0,z_h)/\delta_{{\rm{SFDM}}}(k_0,z_h) = \delta_{{\rm{CDM}}}(k_0,z)/\delta_{{\rm{SFDM}}}(k_0,z)\simeq 1$. Additionally, the overall effect of the last ratio on the right hand side of Eq.~\eqref{critoverdu} occurs for $k > 1$ where $\delta_{\rm{SFDM}}(k,z_h) < \delta_{\rm{CDM}}(k,z_h)$, suppressing the amplitude of the growth factor for such wavenumbers at $z=z_h$, while for $k<1$ such ratio is equal to 1. Besides, notice that $z_h\sim 10
^2$ is the order of magnitude of redshift where the cosmological evolution of the growth factor is basically that of CDM (see Figure~\ref{gfplot}), and thus, the last term in Eq.~\eqref{critoverdu} can be taken as $\delta_{{\rm{CDM}}}(k,z_h)/\delta_{{\rm{SFDM}}}(k,z_h)\simeq 1$, i.e., as almost independent of the wavenumber $k$. Thereby, the main responsible to carry on the scale dependence will be the first ratio in Eq.~\eqref{critoverdu}. We conclude from the above that the critical overdensity can be written as
\begin{equation}
\delta_{crit}(k) =1.686\frac{\delta_{\rm{CDM}}(k,z)}{\delta_{0}(k,z)}\, .
\label{critover}
\end{equation}

From Eq.~\eqref{critover}, it can be seen that for small wavenumbers the CDM case is recovered, since $k\rightarrow 0$ erases the scale-dependence on $\delta_{crit}$. In other words, the density contrast for the axion field will evolve as CDM for small values of $k$, specially at late times. Note that our definition of the critical overdensity given by the above equation is a reduction from that used by authors in \cite{Marsh:2013ywa,Marsh:2016vgj,Du:2016zcv}, where a particular normalization and an analytical function based on axion\textsc{camb} results are implemented for a scale/mass-dependent growth factor. Within our analysis, such scale dependence is encoded in the density contrast given by our new dynamical variable $\delta_0(z,k)$, which we have obtained numerically from \textsc{class}. 

\begin{center}
\begin{figure}[h!]
    \centering
   \includegraphics[width=0.49\linewidth]{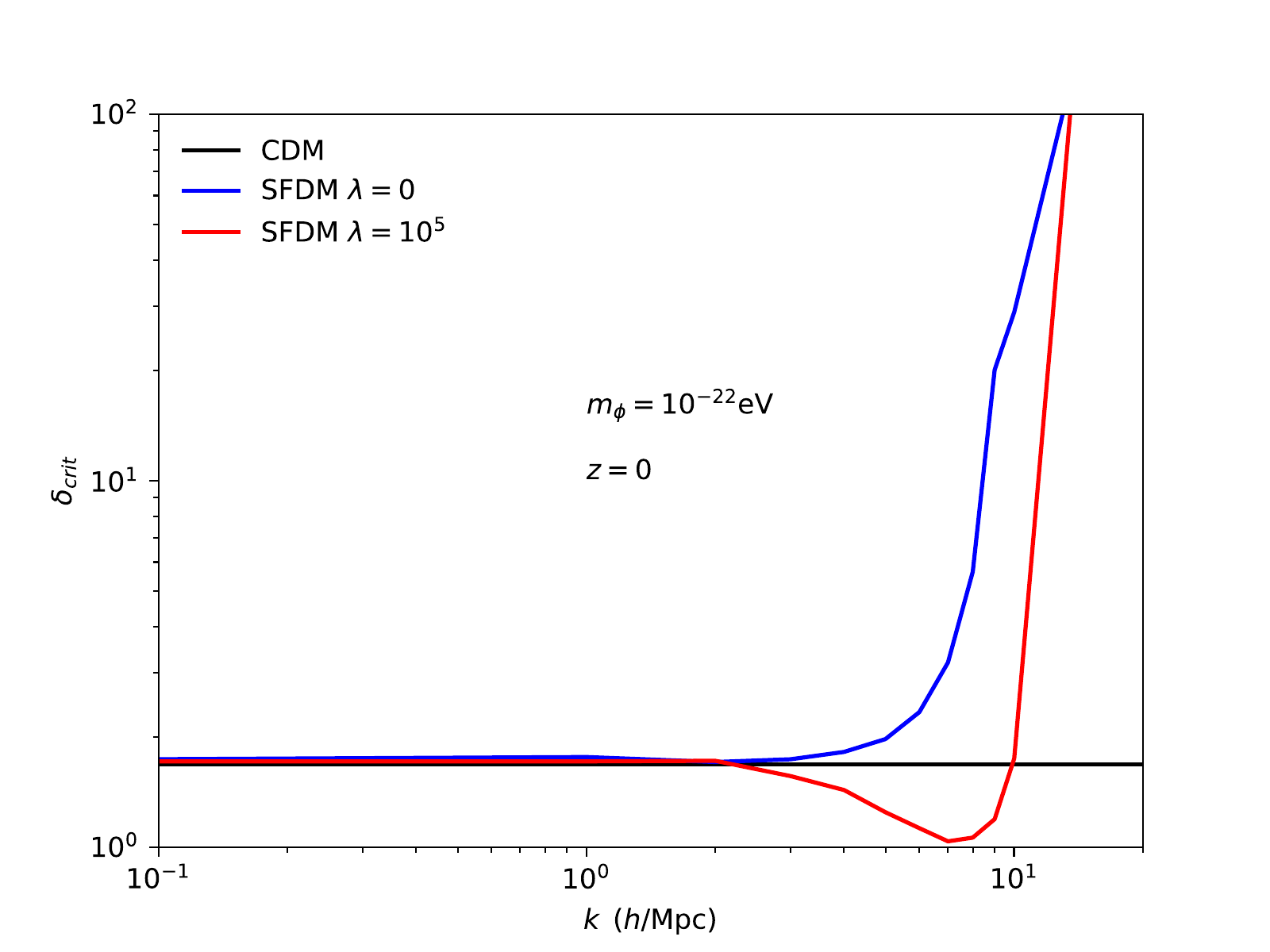}
        \label{fig:first_sub}   
    \caption{Critical overdensity $\delta_{crit}$ at redshift $z=0$ as function of the wavenumber $k$ for an axion field with mass $m_{\phi}=10^{-22}$eV. For an axion with quadratic potential (FDM case, blue line), $ \delta_{crit} $ grows for wavenumbers $ k> 3h / $Mpc, while for the trigonometric potential (red line) there is a decrease for $ 2 < k\ {\rm{Mpc}}/h< 10 $ due to the tachyonic instability, and then grows like the quadratic case. The horizontal black line indicates the value $\delta_{crit} = 1.686$.}
    \label{co}
\end{figure}
\end{center}

We want to highlight that from our definition \eqref{critover} we can recover the results from the previous work mentioned above. For example, Figure~\ref{co} shows the critical overdensity as function of the wavenumber, analogous to that of Figure 2 from \cite{Du:2016zcv}, where $\delta_{crit}$ is shown as function of the mass. Such comparison is valid for an axion mass of $10^{-22}$eV (green curve on Figure 2 from~\cite{Du:2016zcv}, and blue curve in Figure~\ref{co}), since in this work we have consider the effect of tachyonic instability in the critical overdensity as well. We observe that $\delta_{crit}$ shows a clear scale dependence for wavenumbers $k> 1h/$Mpc, which is translated to small halo masses, as we shall see below. Notice that for the trigonometric potential (red curve in Figure~\ref{co}) there are wavenumbers for which the critical overdensity is less than in the CDM case, implying that structures associated to such modes will be able to grow with a threshold $\delta_{crit}$ lower than in a standard CDM scenario, and also compared to the case of a free axion. This is why, from the perspective of the critical overdentisy $\delta_{crit}$, the MPS in Figure \ref{mps} exhibits a bump at small scales.

On modeling the gravitational collapse we will consider both the Press-Schechter (P\&S) and the Sheth-Tormen (S\&T) formalism for spherical and ellipsoidal collapse models~\cite{Press:1973iz,Sheth:1999mn}. Such collapse models are encrypted in the following function,
\begin{equation}
f(\nu) = \left\{
        \begin{array}{ll}
            \sqrt[]{\frac{2\nu}{\pi}}e^{-\nu/2}\, \quad \text{for P\&S}, \\
            A\ \sqrt[]{\frac{2q\nu}{\pi}}(1 + q\nu)^{-p}e^{-q\nu/2}\, \quad \text{for S\&T},
        \end{array}
    \right.
\label{cfm}
\end{equation}
where $\nu \equiv \delta_{crit}^2/\sigma^2$ is the peak height of perturbations, while $A=0.3222\, , p=0.3\, , q=0.707$ for the S\&T model in Eq.~\eqref{cfm} according to \cite{Schneider:2013ria}. Finally, the semi-analytical HMF has the following expression
\begin{equation}
\frac{dn}{d\ln M} = -\frac{1}{2}\frac{\bar{\rho}}{M}f(\nu)\frac{d\ln \sigma^2}{d\ln M}\, .
\label{hmfeq}
\end{equation}

Now we can analyze the HMF for an axion field endowed with a trigonometric potential, and compare it with the CDM prediction, as well as with the free axion case. Figure \ref{hmf} shows the semi-analytical HMF at redshift $z=0$ and axion mass $m_{\phi}=10^{-22}$eV.
\begin{figure}[h!]
\centering
  \includegraphics[width=0.47\linewidth]{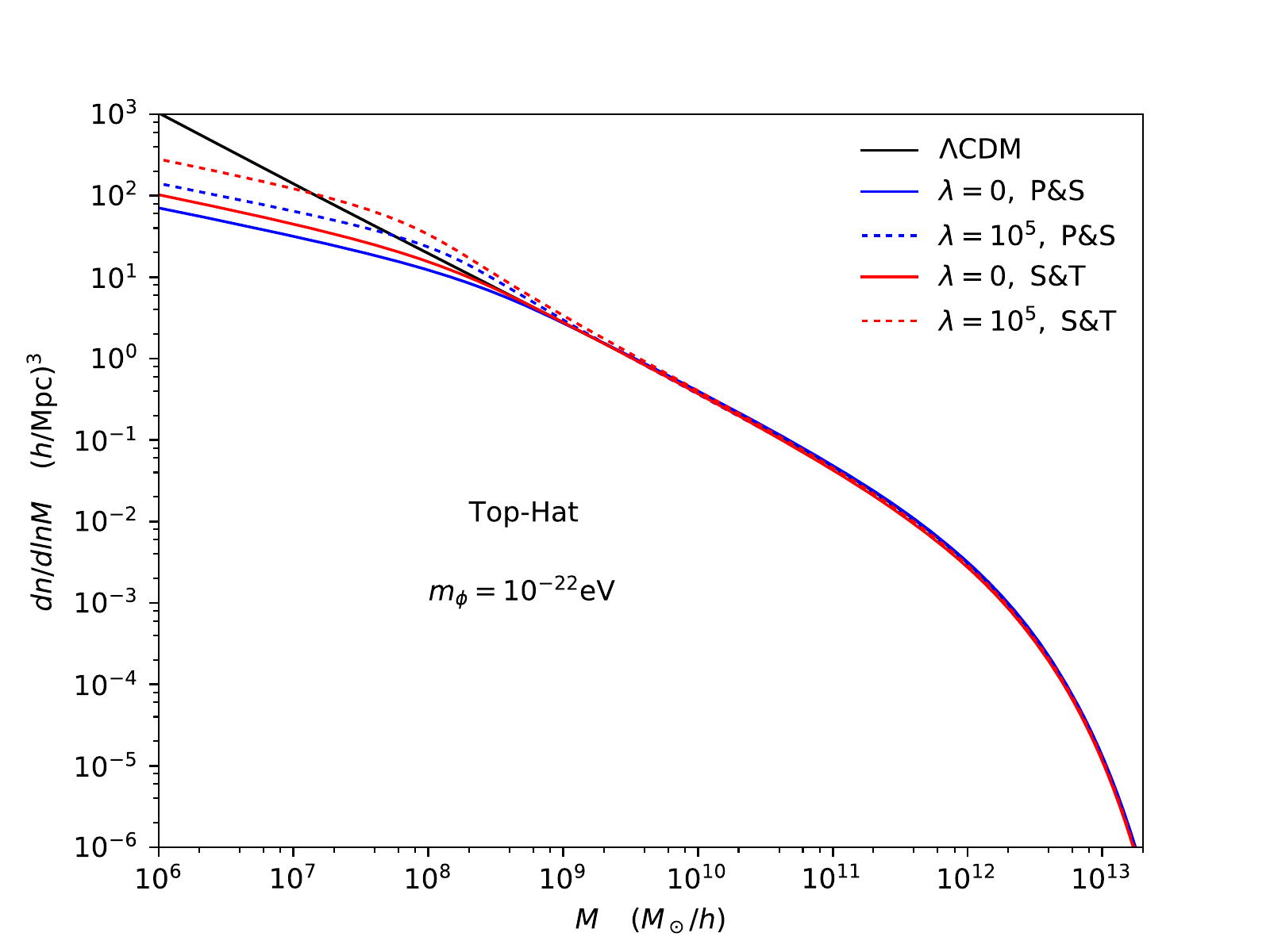}
  \label{fig:evaluation:revenue}
\qquad
  \includegraphics[width=0.47\linewidth]{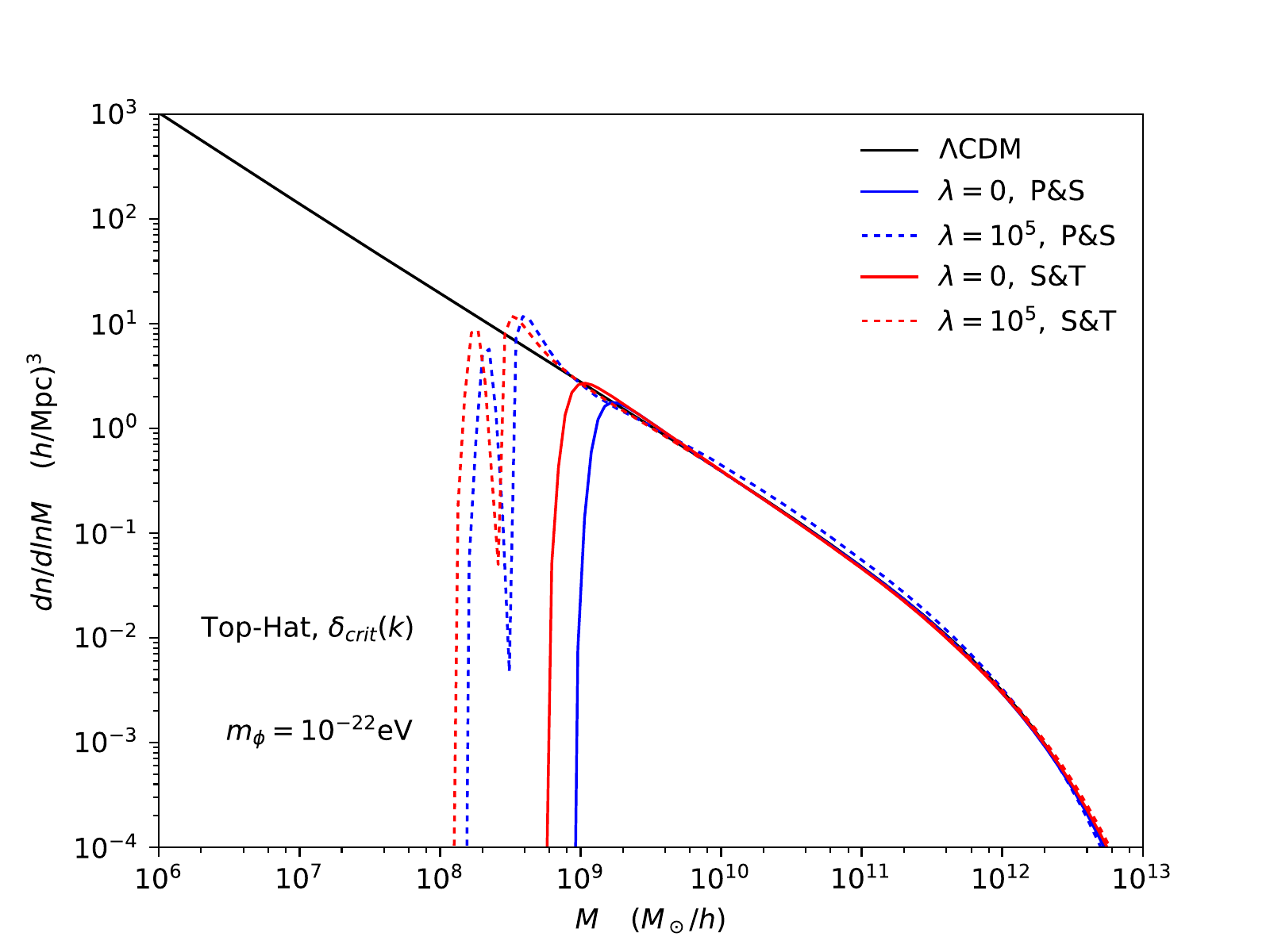}
  \label{fig:evaluation:avgPrice}
\qquad
  \includegraphics[width=0.47\linewidth]{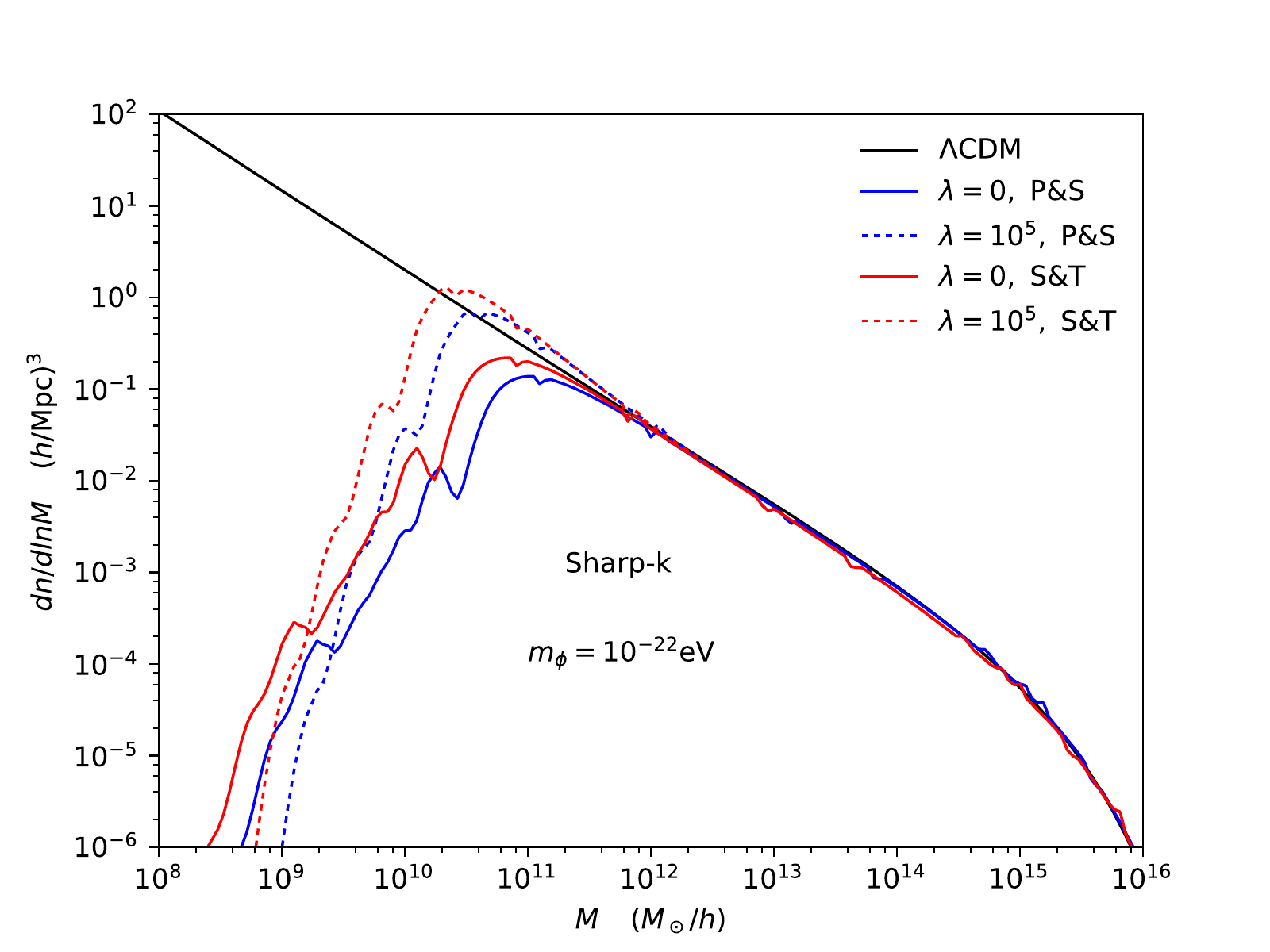}
  \label{hmf}  
\caption{Halo Mass Function for the CDM (black line) and SFDM models with the same values of $\lambda, m_{\phi}$ and $z$ as Figure \ref{sigma}. Two different collapse models are shown, P\&S (blue lines) and S\&T (red lines). Upper left: HMF with a Top Hat window function. Upper right: HMF with a Top Hat window function and scale dependent critical overdensity. Bottom: HMF with a Sharp-k filter.}
\label{hmf}
\end{figure}

We separate our analysis in three different cases depending on the window function implemented: Top-Hat, Sharp-k, and Top-Hat including the critical overdensity with scale dependence. For all cases we consider the collapse models given at Eq.~\eqref{cfm}, for the FDM case with quadratic potential ($\lambda=0$) and an axion field with trigonometric potential ($\lambda=10^5$). When considering the Top Hat window function $W_{TH}$ without a scale dependent critical overdensity, differences between the HMF for SFDM and CDM appear at small mass scales, as can be seen at upper left panel in Figure \ref{hmf}. However, since we have used Eq.~\eqref{critoversfdm}, the HMF do not exhibits the cut-off of the MPS when using this window function. This is because, as we mentioned before, the dependence on $k$ in the growth factor \eqref{gf} is lost at late times. On the other hand, including the scale-dependent critical overdensity~\eqref{critover} (upper right panel in Figure \ref{hmf}), a steep cut-off appears at $M \sim 10^9 M_\odot/h$ for $\lambda=0$ and $M \sim 10^8 M_\odot/h$ for $\lambda=10^5$. This result is consistent with that of \cite{Marsh:2013ywa,Du:2016zcv} for the particular case in which SFDM constitutes all the DM content, i.e., when $\Omega_{\phi}/\Omega_{\rm{CDM}} = 1$. Finally, the HMF with the Sharp-k function $W_{SK}$ is shown in the lower panel of Figure \ref{hmf}. In this case, we have used Eq.~\eqref{critoversfdm}, since the cut-off at a given scale is captured by the Sharp-k window function, as discussed by \cite{Urena-Lopez:2015gur}. Although the turn around of the HMF is slightly different for $\lambda=0$ and $\lambda=10^5$, the cut-off for both of them occurs approximately at the same range of mass scale $10^{8} \lesssim M \ \left(h/M_{\odot}\right)\lesssim 10^9$.

For all the cases studied, we observe as a new general feature in the HMF an increment in its amplitude when considering one of the two following considerations:
\begin{itemize}
    \item[1.-] Ellipsoidal collapse S\&T model (red lines in Figure~\ref{hmf}),
    \item[2.-] Axion-like potential in the tachyonic instability regime (dotted lines in Figure~\ref{hmf}).
\end{itemize}

At this point, it would be interesting to contrast the result of such considerations either to data, or to simulations. We will take the former approach and use data from stellar streams measurements, particularly the GD-1 stream detected in Sloan Digital Sky Survey (SDSS) data~\cite{Grillmair:2006bd,Banik:2019cza,Banik:2019smi}. Stellar streams are originated from the tidal disruption of globular clusters, forming an elongated structure that, when gravitationally perturbed by dark subhaloes, some gaps in the stellar distribution of such elongated structure are produced. Therefore, stellar stream observations would provide information about the DM subhaloes~\cite{Ibata:2001iv,Yoon:2010iy,Carlberg:2011xj,Bovy:2016irg,erkal2016number,Bonaca:2018fek,Banik:2019smi,Benito:2020avv}.
So we will use this data and compare constraints using our modelling of the HMF with those obtained by \cite{Schutz:2020jox} for the FDM model.

\begin{center}
\begin{figure}[h!]
    \centering
   \includegraphics[width=0.7\linewidth]{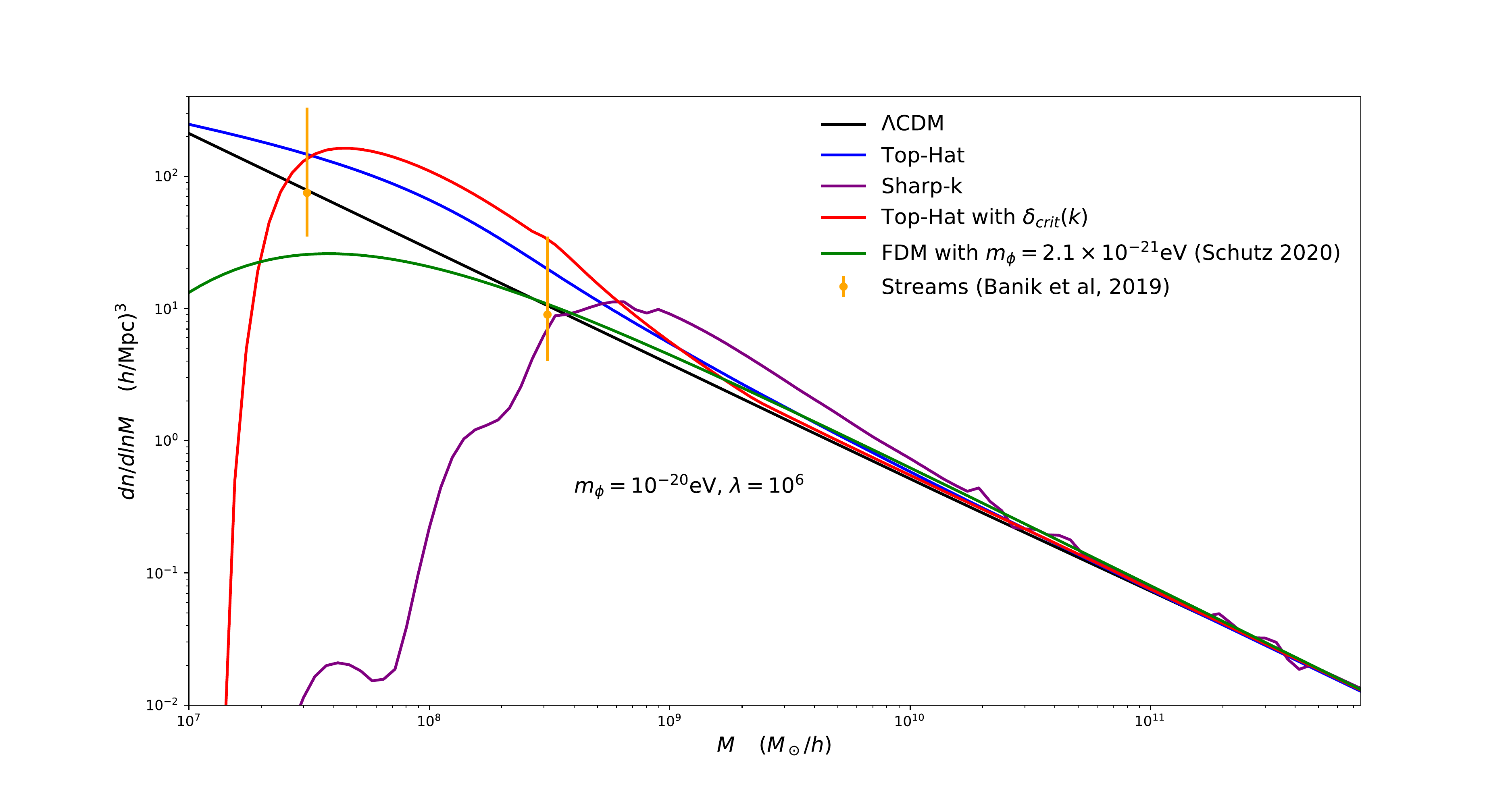}
        \label{fig:first_sub}   
    \caption{HMF for the analytical approach by~\cite{Schutz:2020jox} for FDM with $m_{\phi}=2.1\times 10^{-21}$eV (green), and our numerical results for SFDM HMF with $m_{\phi}=10^{-20}$eV, $\lambda = 10^{6}$, for a Top-Hat (blue), Sharp-k (purple), and Top-Hat window function with scale-dependent critical overdensity (red). All of them considering the S\&T collapse model. Orange error bars show data from streams measurements~\cite{Banik:2019cza,Banik:2019smi}. See text for more details.}
    \label{hmf_comp}
\end{figure}
\end{center}

We show in Figure~\ref{hmf_comp} the HMF obtained by~\cite{Schutz:2020jox} when considering $m_{\phi} = 2.1\times 10^{-21}$eV (green solid line), as well as the corresponding HMF for the CDM model (black solid line). We have also included the numerical results for the HMF obtained in this work for an axion mass of  $m_{\phi} = 10^{-20}$eV and $\lambda = 10^{6}$ (blue, purple and red lines). Stellar streams measurements from~\cite{Banik:2019cza,Banik:2019smi} are included as well (orange points). 
Our numerical results for the HMF using the Sharp-k and Top-hat with scale dependent critical overdensity with axion mass $m_{\phi}<10^{-20}$ fails to reproduce the observations regardless the value of $\lambda$. Particularly for the value of $m_{\phi}=2.1\times 10^{-21}$ the HMF falls below the analytical approach used in~\cite{Schutz:2020jox} (green line). Using the simple Top-Hat window function allows for lighter mass values and agrees better with~\cite{Schutz:2020jox}. However, as discussed before, this window function is not well motivated from the simulations point of view for models with suppression in the MPS~\cite{Schneider:2014rda}.

On the other hand, we found that for $m_{\phi} \geq 10^{-20}$eV there is variety of combinations of $m_{\phi}$ and $\lambda$ that results in a HMF in agreement with streams measurements when both, the Top-Hat with a scale-dependent critical overdensity and the S\&T collapse model are considered. In particular, in Figure~\ref{hmf_comp} (red curve) we show the case of  $m_{\phi}=10^{-20}$ and $\lambda=10^6$. This leads to stronger constraints as those imposed by Lyman-$\alpha$~\cite{Irsic:2017yje,Armengaud:2017nkf}, but lies within the range of masses that could be tested by 21-cm observations~\cite{Munoz:2019hjh,Shimabukuro:2019gzu}.


It can also be seen that the HFM for a Sharp-k window function (purple curve) requires larger masses to be in agreement with both measurements of stellar streams. Therefore, even when both window functions contain information of the MPS suppression at small scales, it seems that the Top-Hat with a scale-dependent critical overdensity is less restrictive than that of a Sharp-k window function, in the sense that the latter requires $m_{\phi}>10^{-20}$eV.  Nonetheless, to find out which of the different window functions gives the more realistic HMF one has to compare and calibrate the methods using numerical simulations of structure formation specific for the SFDM model. Specifically, recall that the Sharp-k function has one free parameter, $c$, that is tuned using simulations. It would be possible that this parameter could be tuned using both, data and simulations.

As it is noted in~\cite{Schutz:2020jox}, the results from the analytical approach are too conservative in the sense that they are not taking into account the scale-dependent growth of structure, neither the scale-dependent critical overdensity. In their analysis, masses for FDM with values $m_{\phi} \lesssim 1.37\times 10^{-20}$eV would be excluded, whereas we are showing that the SFDM HMF (blue and red curves) lie within the range obtained from stellar streams measurements for $m_{\phi} = 10^{-20}$eV. This value is consistent, however, with the conservative bound where masses $m_{\phi}\lesssim 2.1\times 10^{21}$eV are excluded. Without the effect of the tachyonic instability, stronger constraints on the axion mass would be imposed.

\section{Conclusions}\label{sec:conclusions}
The SFDM model constitutes a compelling candidate to substitute the CDM model. In this paper, we have presented a formalism to handle the cosmological equations by using the tools of dynamical systems for both the background and the linear perturbations, that extends the analysis in~\cite{Cedeno:2017sou}. At the background level, the presence of a trigonometric potential shows a delay in the moment when the axion field starts to oscillate and behaving as CDM. These values of the onset of oscillations are shown in Table \ref{table} in terms of the scale factor $a_{\rm{osc}}$. We have explored with some depth the effect dubbed as tachyonic instability, which occurs due to the non--linearities of the potential \eqref{eq:0}. For extreme values of $\lambda$, once the axion field starts to oscillate the density contrast grows with more amplitude than that of standard CDM and SFDM with a quadratic potential (FDM). We indicated the duration of the tachyonic instability as well as the range of wavenumbers that suffer such effect. This is the particular imprint of the axion--like potential~\eqref{eq:0} on the formation of large scale structures.

When performing the statistical analysis with data from Planck Collaboration 2018, we obtain a lower bound for the axion mass given by $m_{\phi} \gtrsim 10^{-24}$eV at $95.5\%$ C.L. By including Lyman-$\alpha$ observations, a stronger constraint is imposed on the axion mass given by $m_{\phi} \gtrsim 10^{-22}$eV at $95.5\%$ C.L. This is important to highlight, as we are showing that once the scalar field is endowed with the true axion potential the fiducial mass $m_{\phi}=10^{-22}$eV is again consistent with these observations. The decay parameter $\lambda$ is not constrained, and all the values we explored are numerically limited to a particular maximum value $\lambda_{max}$ for any given axion mass. As stressed in the text, the constraints discussed above are valid for an axion field providing all the DM component in the universe.

Motivated by the scale-dependence of the scalar field dark matter models, we proposed a growth factor $D_k$ and a velocity growth factor $f_k$ with explicit dependence on the wavenumber $k$. This was performed through the density contrast and its derivative. Effectively, there are differences in the evolution of $D_k$ and $f_k$ for each value of $k$, but all of them evolve as cold DM from certain value of redshift ($z\sim 100$ for $D_k$, and $z\sim 10$ for $f_k$) up to the present day ($z \sim 0$). Having this quantities allowed us to build the combined parameter $f_k\sigma_8$, which shows marginal differences when compared to CDM.

The tachyonic instability of the axion potential is manifested in both the variance and the HMF as an enhancement in the amplitude for low halo masses. The HMF with a Top Hat window function presents a decrease at small halo masses for SFDM in comparison with the CDM model, but such decrease is not the one expected from a MPS with a cut-off. However, when considering a scale-dependent critical overdensity, the HMF exhibits a steep cut-off. With the Sharp-k window function the HMF has a cut-off less pronounced than the mentioned above, but it still appears at approximately the same mass scales. All the cases were studied for two different gravitational collapse models, the Press-Schechter and the Sheth-Tormen for the spherical and ellipsoidal collapse, respectively. Both of them produce qualitatively the same HMF, with small differences at small scales. In particular, stellar stream measurements seem to indicate that $m_{\phi}\gtrsim 10^{-20}$eV, which is a stronger bound than that obtained from Lyman-$\alpha$ observations.

Future astronomical observations planned by collaborations such as the Dark Energy Spectroscopy Instrument (DESI)~\cite{levi2013desi} and the Large Synoptic Survey Telescope (LSST, now Vera C. Rubin Observatory)~\cite{Ivezic:2008fe} will explore the Universe with a better accuracy. Particularly, the LSST will be able to constraint the SFDM mass $m_{\phi}\sim 10^{-20}$eV by probing the MPS for halos with $\sim 10^8 M_{\odot}$~\cite{Drlica-Wagner:2019xan}. On the other hand, the Sloan Digital Sky Survey (SDDS) can be used for searches of low-surface brightness dwarf galaxies at small scales, as discussed in ~\cite{Muller:2017ehk}. Besides, the 21cm signal detected by EDGES~\cite{Bowman:2018yin} can be used to study the properties of DM~\cite{Barkana:2018lgd}, in particular to probe small scale structures~\cite{Munoz:2019hjh}. In fact, forthcoming experiments such as the Square Kilometre Array (SKA)~\cite{Bull:2018lat}, 21cm observations can be used to constraint the scalar field mass as well at wavenumbers $30<k<1000\ $({\rm{Mpc}}$^{-1})$~\cite{Shimabukuro:2019gzu}. There may be implications of a post-inflationary symmetry breaking of axion-like particles in the formation of the first generation of stars, and on small scale structures~\cite{Irsic:2019iff}. For a recent review on other gravitational probes for ultra-light axions see~\cite{Grin:2019mub}. Therefore, the physics at these small scales and high redshifts will in the near future reveal more information about the properties of axion-like DM. 

\acknowledgments
Francisco X. Linares Cedeño acknowledges the receipt of the grant from the Abdus Salam International Centre for Theoretical Physics, Trieste, Italy. FXLC also acknowledges CONACYT and the Programa para el Desarrollo Profesional Docente for financial support.  This work was partially supported by Programa para el Desarrollo Profesional Docente; Direcci\'on de Apoyo a la Investigaci\'on y al Posgrado, Universidad de Guanajuato under Grant No. 099/2020; CONACyT M\'exico under Grants No. A1-S-17899, 286897, 297771, 304001; and the Instituto Avanzado de Cosmología collaboration.

\appendix

\section{Higher order algebraic equation for the scale factor on the onset of oscillations}
\label{sec:higher-order}
In Section~\ref{blp}, we obtained an expression to determine the scale factor at the onset of oscillation $a_{\rm{osc}}$ given by Eq.~\eqref{cqe}, which was important to determine the initial conditions for the evolution of the background variables when $\lambda > 0$. Here we will show that it is possible to obtain a more accurate expression for $a_{\rm{osc}}$ by means of an iterative integration of the equations of motion at early times. 

Considering again a radiation domination era, let us take the solution for $y_1$ given by Eq.~\eqref{y1fo} and plug it into Eq.~\eqref{eq:new4b}, which leads to the new solution,
\begin{equation}
y_1(a) = y_{1,i}\left( \frac{a}{a_i} \right)^{2} + \frac{\lambda}{8}\Omega_{\phi,i}\theta_i \left( \frac{a}{a_i} \right)^{6} - \left( \frac{\lambda \Omega_{\phi,i}}{24} \right)^2\theta_i \left( \frac{a}{a_i} \right)^{6} + \frac{1}{2}\left( \frac{\lambda \Omega_{\phi,i}}{24} \right)^2\theta_i \left( \frac{a}{a_i} \right)^{10}\, .
\label{y1so}
\end{equation}
This solution can be used in Eq.~\eqref{eq:new4a} to obtain the following new solution for $\theta$,
\begin{equation}
\theta(a) = \theta_i \left( \frac{a}{a_i} \right)^{2}\left\lbrace \left[ 1 - \frac{\lambda \Omega_{\phi,i}}{72} + \frac{17}{26}\left( \frac{\lambda \Omega_{\phi,i}}{72} \right)^2 \right] + \frac{\lambda \Omega_{\phi,i}}{72}\left( 1 - \frac{\lambda \Omega_{\phi,i}}{72} \right)\left( \frac{a}{a_i} \right)^{4} + \frac{9}{26}\left( \frac{\lambda \Omega_{\phi,i}}{72} \right)^2 \left( \frac{a}{a_i} \right)^{8} \right\rbrace\, .
\label{tso}
\end{equation}
Setting the previous expression to the onset of oscillations, $a=a_{\rm{osc}}$ and $\theta = \pi/2$, we obtain a quartic order equation for $a_{\rm{osc}}$,
\begin{equation}
a_{\rm{osc}}^2\left[ 1 + \left(\frac{\lambda}{72}\frac{\Omega_{\phi,0}}{\Omega_{r,0}}\right)a_{\rm{osc}} + \frac{9}{26}\left( \frac{\lambda}{72}\frac{\Omega_{\phi,0}}{\Omega_{r,0}}\right)^2 a_{\rm{osc}}^2 \right] = \frac{\pi \theta^{-1}_i a_i^2}{2\ \sqrt[]{1 + \pi^2/36}}\, .
\label{aosc4}
\end{equation}

Following the same iterative scheme, we can find higher order solutions for $y_1$ and $\theta$, which we do not show, but that lead to a fifth order equation for $a_{\rm{osc}}$,
\begin{equation}
a_{\rm{osc}}^2\left[ 1 + \left(\frac{\lambda}{72}\frac{\Omega_{\phi,0}}{\Omega_{r,0}}\right)a_{\rm{osc}} + \frac{9}{26}\left( \frac{\lambda}{72}\frac{\Omega_{\phi,0}}{\Omega_{r,0}}\right)^2 a_{\rm{osc}}^2 + \frac{27}{442}\left( \frac{\lambda}{72}\frac{\Omega_{\phi,0}}{\Omega_{r,0}}\right)^3 a_{\rm{osc}}^3\right] = \frac{\pi \theta^{-1}_i a_i^2}{2\ \sqrt[]{1 + \pi^2/36}}\, .
\label{aosc5}
\end{equation}

Higher order solutions follow a similar pattern as that in Eq.~\eqref{aosc5}, which resembles the series expansion of the exponential series, except for the numerical coefficients. But from a close comparison between the two series we find that,
\begin{eqnarray}
&& 1 + \left(\frac{\lambda}{72}\frac{\Omega_{\phi,0}}{\Omega_{r,0}}\right)a_{\rm{osc}} + \frac{9}{26}\left( \frac{\lambda}{72}\frac{\Omega_{\phi,0}}{\Omega_{r,0}}\right)^2 a_{\rm{osc}}^2 + \frac{27}{442}\left( \frac{\lambda}{72}\frac{\Omega_{\phi,0}}{\Omega_{r,0}}\right)^3 a_{\rm{osc}}^3 + \cdots  \nonumber \\
< \ && 1 + \left(\frac{\lambda}{72}\frac{\Omega_{\phi,0}}{\Omega_{r,0}}\right)a_{\rm{osc}} + \frac{1}{2}\left( \frac{\lambda}{72}\frac{\Omega_{\phi,0}}{\Omega_{r,0}}\right)^2 a_{\rm{osc}}^2 + \frac{1}{6}\left( \frac{\lambda}{72}\frac{\Omega_{\phi,0}}{\Omega_{r,0}}\right)^3 a_{\rm{osc}}^3 + \cdots =e^{\left(\frac{\lambda}{72}\frac{\Omega_{\phi,0}}{\Omega_{r,0}}\right)a_{\rm{osc}}}\, . \nonumber \\
\label{expaosc}
\end{eqnarray}
Although not a formal demonstration, this exercise shows that a better estimation of the scale factor at the onset of the oscillations could be made from the expression
\begin{equation}
a^2_{\rm{osc}} \exp \left( \frac{\lambda}{72}\frac{\Omega_{\phi,0}}{\Omega_{r,0}} a_{\rm{osc}} \right) = \frac{\pi \theta^{-1}_i a_i^2}{2\ \sqrt[]{1 + \pi^2/36}} \, . \label{eq:exponential}
\end{equation}

As discussed in Section~\ref{blp}, the start of the field oscillations happen more abruptly for larger values of $\lambda$, and this makes difficult to find the right initial conditions for the dynamical variables. Equation~\eqref{eq:exponential} seems to offer an explanation of this, as the iterative integration of the equations of motion results in an (nearly) exponential relationship between $a_{\rm osc}$ and the values of other cosmological variables.

\section{Extreme axion Wave Dark Matter \label{sec:ewdm}}
The tachyonic instability of SFDM in the axion case was firstly studied in~\cite{Chiueh:2014qla}, from the field perspective, and was dubbed Extreme axion Wave Dark Matter (EA$\psi$DM). Assuming an axion potential in the form $V(\phi) = 2 m^2_\phi f^2_\phi \sin^2(\phi/2f_\phi)$, the dynamics of the field starts close to maximum of the potential, and then the extreme label refers to initial conditions such that $\phi_i/f_\phi \to \pi$. For instance, some of the most extreme values considered in~\cite{Zhang:2017flu} were of the order $\delta \theta_0 \equiv \pi - \phi_i/f_\phi \simeq 0.2^\circ$.

To find the relation between the extreme initial conditions used in~\cite{Zhang:2017flu} and our approach, we proceed as follows. Considering our convention for the axion potential~\eqref{eq:trigonometric}, we find for the initial conditions that
\begin{equation}
    \frac{2m^2_\phi}{H^2_0} \frac{a^4_i}{\Omega_{r0} \lambda} \cos (\phi_i/2f_\phi) = \Omega_{\phi i} \, . \label{eq:missal}
\end{equation}
In our convention, EA$\psi$DM is achieved if $\phi_i /f_\phi \to 0$, and then we see that an extreme initial condition on the field $\phi_i$ translates into an extreme initial condition on the density parameter $\Omega_{\phi i} \to 0$. However, the latter's value is not independent, as for any choice of the potential parameters $m_\phi$ and $\lambda$ (i.e., $f_\phi$), one has to fine tune $\Omega_{\phi i}$ to get the right value of $\Omega_{\phi 0}$ at the present time.

The above is the main reason why, in our approach, the extreme case of initial conditions is interlinked with the (decay) parameter $\lambda$: larger values of the latter asks for smaller values of $\phi_i/f_\phi$, that is, for more extreme values in the sense that $\phi_i/f_\phi \to 0$. For the fiducial model with $m_\phi = 10^{-22}$, we find $\delta \theta_0 = (174^\circ,162^\circ,124^\circ,44^\circ,0.47^\circ,0.16^\circ)$ corresponding to $\lambda = (10,10^2,10^3,10^4,10^5,1.28\times 10^5)$. Thus, our formalism allows initial conditions as extreme as those reported in~\cite{Zhang:2017flu}, but covering the whole evolution of the Universe.

As a side result, we show in Table~\ref{tab:extreme} the pairs of values $(m_\phi,\lambda_{max})$ and their corresponding missalignment $\delta \theta_0$, as calculated from Eq.~\eqref{eq:missal}. Here, $\lambda_{max}$ is the maximum value of the parameter $\lambda$ allowed by the numerical accuracy of the amended version of \texttt{CLASS} for a given mass $m_\phi$. We can see that there is simple relation between $m_\phi$ and $\lambda_{max}$: their values are in correspondence with the smallest displacement of the field with respect to the maximum in the potential~\eqref{eq:0}, which is $\delta \theta_0 \simeq 0.16^\circ$ (see also Figure~\ref{cp}).

\begin{table}[h!]
\centering
\begin{tabular}{|c|c|c|c|c|c|c|}
\hline
$m_\phi$ (eV) & $10^{-26}$ & $10^{-25}$ & $10^{-24}$ & $10^{-23}$ & $10^{-22}$ & $10^{-21}$ \\ 
\hline 
$\lambda_{max}$ & $1.45 \times 10^3$ & $4.22 \times 10^3$ & $1.29 \times 10^4$ & $4.05 \times 10^4$ & $1.28 \times 10^5$ & $4.03 \times 10^5$ \\ 
\hline 
$\delta \theta_0$ & $0.162^\circ$ & $0.163^\circ$ & $0.166^\circ$ &$0.163^\circ$ & $0.162^\circ$ & $0.164^\circ$ \\ 
\hline
\hline
$m_\phi$ (eV) & $10^{-20}$ & $10^{-19}$ & $10^{-18}$ & $10^{-17}$ & $10^{-16}$ & $10^{-15}$ \\ \hline 
$\lambda_{max}$ & $1.27 \times 10^6$ & $4.04 \times 10^6$ & $1.27 \times 10^7$ & $4.03 \times 10^7$ & $1.27 \times 10^8$ & $4.04 \times 10^8$ \\ \hline 
$\delta \theta_0$ & $0.168^\circ$ & $0.164^\circ$ & $0.167^\circ$ &$0.164^\circ$ & $0.170^\circ$ & $0.164^\circ$ \\ \hline 
\end{tabular}
\caption{\label{tab:extreme} The field mass $m_\phi$ and its corresponding extreme value $\lambda_{max}$. In the last row we show the initial field displacement from the top of the axion potential $\delta \theta_0$, for a comparison with the EA$\psi$DM model in~\cite{Zhang:2017flu}.}
\label{table}
\end{table}

\section{General dynamical variables for scalar field perturbations}\label{sfp}
To translate the same scheme we used for the background evolution of axion fields in Section~\ref{blp}, where we were able to write down a dynamical system for the KG equation, we propose the following new variables for the scalar field perturbation $\varphi$ and its derivative $\dot{\varphi}$~\cite{Urena-Lopez:2015gur},
\begin{equation}
u = \sqrt{\frac{2}{3}} \frac{\kappa \dot{\varphi}}{H} = -\Omega^{1/2}_{\phi}e^{\alpha}\cos(\vartheta/2) \, , \label{eq:22a} \quad
    v = \frac{\kappa y_1 \varphi}{\sqrt{6}} = -\Omega^{1/2}_{\phi}e^{\alpha}\sin(\vartheta/2) \, ,
\end{equation}
which after substitution on the perturbed KG equation~\eqref{eq:13} lead to the following differential equations
\begin{subequations}
\label{pertvardiffeq}
\begin{eqnarray}
\vartheta' &=& 3\sin \vartheta +2\omega \left( 1-\cos \vartheta \right)+y_1-2e^{-\alpha}h'\sin \left( \frac{\theta}{2} \right)\sin \left( \frac{\vartheta}{2} \right) \nonumber \\
&& + \frac{\lambda \Omega_\phi}{y_1} \cos\left( \frac{\theta}{2} \right) \left[ \cos\left( \vartheta - \frac{\theta}{2} \right) - \cos\left( \frac{\theta}{2} \right) \right] \, , \label{vartheta}\\
\alpha' &=& -\frac{3}{2}\left( \cos \vartheta + \cos \theta \right)-\omega\sin \vartheta+e^{-\alpha}h'\sin\left( \frac{\theta}{2} \right)\cos\left( \frac{\vartheta}{2} \right) \nonumber \\
&& + \frac{\lambda \Omega_\phi}{2y_1} \left[\sin\left( \frac{\theta}{2} \right) + \sin\left(\vartheta -  \frac{\theta}{2} \right)  \right] \, .
\end{eqnarray}
\end{subequations}
For numerical purposes, it is convenient to use as angular variable the difference $\tilde{\vartheta}\equiv \theta - \vartheta$. If we further define the variables $\delta_0=-e^{\alpha}\sin(\tilde{\vartheta}/2)$ and $\delta_1=-e^{\alpha}\cos(\tilde{\vartheta}/2)$, Eq.~\eqref{pertvardiffeq} can be properly combined to obtain the dynamical system shown in Eq.~\eqref{eqdeltas}.

\section{Fluid interpretation of SFDM density perturbations}
\label{sec:eawdm}
Here we report about the fluid interpretation of Eq.~\eqref{eqnewdeltas} in terms of the standard fluid variables for linear perturbations, namely the density contrast $\delta_{\phi} = \delta_0$ and the divergence of the velocity perturbation $\theta_\phi$ (see Eq.~\eqref{eq:26c}). Following the procedure in~\cite{Cookmeyer:2019rna}, we first consider the equations of density perturbations well within the regime of rapid field oscillations, Eqs.~\eqref{eqnewdeltas1}, but written in the form,
\begin{equation}
    \delta^\prime_0 =  - \theta_\phi - \frac{\bar{h}^\prime}{2} \, ,\quad \theta^\prime_\phi = - \frac{a^\prime}{a} \theta_\phi + \frac{k^4}{4 a^2 m^2_\phi} \left( 1- \frac{\rho_\phi a^2}{2 k^2 f^2_\phi} \right) \delta_0 \, ,
\end{equation}
where now a prime denotes derivative with respect to $\tau$. Notice that we have used the relation $\theta_\phi = \frac{k^2}{2am_\phi} \delta_1$, which is found from Eq.~\eqref{eq:26c} for rapid oscillations. A quick comparison with the standard fluid equations for axion fields (see for instance Eqs.~(13) and~(14) in~\cite{Cookmeyer:2019rna}), leads us to conclude that the averaged value of the sound speed $c_s$ of the axion field, in the nonrelativistic limit, is given by
\begin{equation}
    \langle c^2_s \rangle \simeq \frac{k^4}{4 a^2 m^2_\phi} \left( 1- \frac{\rho_\phi a^2}{2 k^2 f^2_\phi} \right) \, .
\end{equation}
The standard result of the FDM case is obtained in the limit $f_\phi \to \infty$ ($\lambda \to 0$), namely $\langle c^2_s \rangle_0 \simeq \frac{k^4}{4 a^2 m^2_\phi}$ (eg~\cite{Park:2012ru}).

\section{Analytical transfer function $T(k)$}\label{abg}
The latest version of \textsc{Monte Python} includes a likelihood from Lyman-$\alpha$ observations to constrain non--cold DM models at small scales, which is based on an analytical expression for the transfer function proposed in~\cite{Murgia:2017lwo,Murgia:2017cvj,Murgia:2018now}. Inspired by the transfer function of warm DM models, the transfer function is given by the expression,
\begin{equation}
    T(k) = \left[ 1 + \left( \alpha k \right)^{\beta} \right]^{\gamma}\, ,
    \label{tf}
\end{equation}
where $\alpha$ sets the scale of suppression of structure formation, and $\beta$ and $\gamma$ set the shape of the cut--off.

All the models parameterized by Eq.~\eqref{tf} show a cut--off at small scales that decreases monotonically with the wavenumber (see for instance Fig.1 in~\cite{Murgia:2018now}), and therefore, as far as the axion field with a trigonometric potential is concerned, it will be not possible for such transfer function to reproduce the characteristic bump at the cut--off scale due to the tachyonic instability. Rather than Eq.~\eqref{tf}, a more appropriate expression would be,
\begin{equation}
    T(k) = \left[ 1 + \left( \alpha k \right)^{\beta} \right]^{\gamma}\times \left( 1 + \delta k^{\eta}\right)^{\nu}\, ,
    \label{mod_tf}
\end{equation}
where $\delta\, , \eta\, , \nu$ are three new parameters controlling the amplitude and cut--off of the bump. 

\begin{center}
\begin{figure}[h!]
    \centering
   \includegraphics[width=0.7\linewidth]{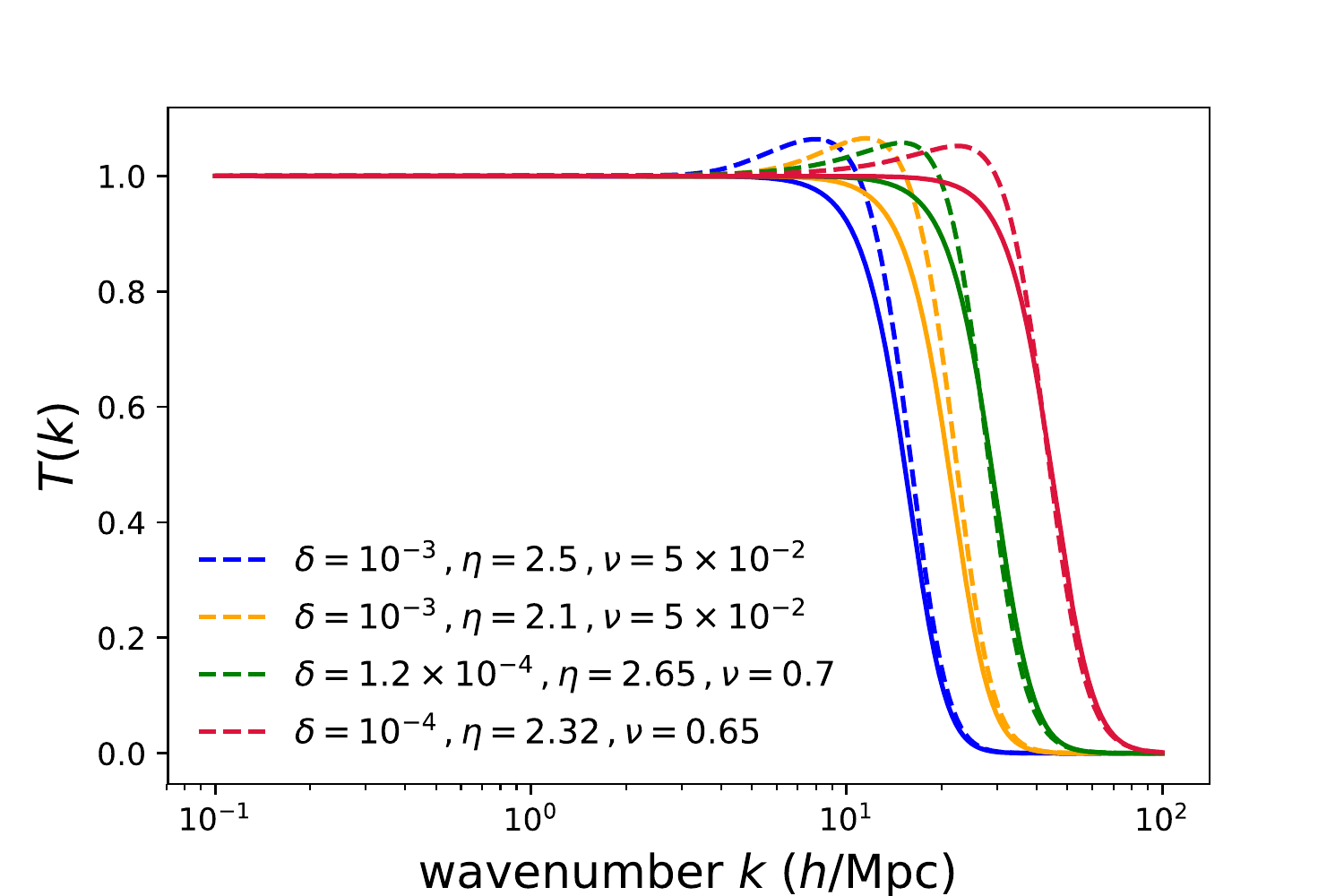}
        \label{fig:first_sub}   
    \caption{Transfer function for FDM according to Eq.~\eqref{tf} (solid lines) and for our model with tachyonic instability given by Eq.~\eqref{mod_tf} (dashed lines). For both transfer functions, the parameters $\alpha\, , \beta\, , \gamma$ have the same values as those for FDM shown in Fig.6 of~\cite{Murgia:2017lwo}. The value of the axion mass $m_{\phi}$ as well as the parameters $\left\lbrace \alpha\, , \beta\, , \gamma \right\rbrace$ are given in Eq.~\eqref{setup}.}
    \label{atf}
\end{figure}
\end{center}

In Figure~\ref{atf} we illustrate the effect of the new modification on $T(k)$. Indicated with solid lines, we have reproduced Figure~6 (right) of~\cite{Murgia:2017lwo}, considering the same values of the FDM mass $m_{22}\equiv m_{\phi}/(10^{-22}{\rm{eV}})$ and their corresponding $\left\lbrace \alpha\, , \beta\, , \gamma \right\rbrace$ parameters,
\begin{subequations}\label{setup}
\begin{eqnarray}
m_{22} &=& 5\, , \  \alpha = 0.054\, , \beta = 5.4\, , \gamma = -2.3\, ,\quad {\rm{(blue).}}  \\
m_{22} &=& 10\, , \alpha = 0.040\, , \beta = 5.4\, , \gamma = -2.1\, ,\quad {\rm{(orange).}}  \\
m_{22} &=& 20\, , \alpha = 0.030\, , \beta = 5.5\, , \gamma = -1.9\, ,\quad {\rm{(green).}}  \\
m_{22} &=& 40\, , \alpha = 0.022\, , \beta = 5.6\, , \gamma = -1.7\, ,\quad {\rm{(red).}} 
\end{eqnarray}
\end{subequations}

The modified transfer function~\eqref{mod_tf} (dashed lines) have the same $\left\lbrace \alpha\, , \beta\, , \gamma \right\rbrace$ values as the FDM case. It can be seen that the new term that we have included in the transfer function is able to reproduce the characteristic bump at the cut--off scale.

\begin{center}
\begin{figure}[h!]
    \centering
   \includegraphics[width=0.7\linewidth]{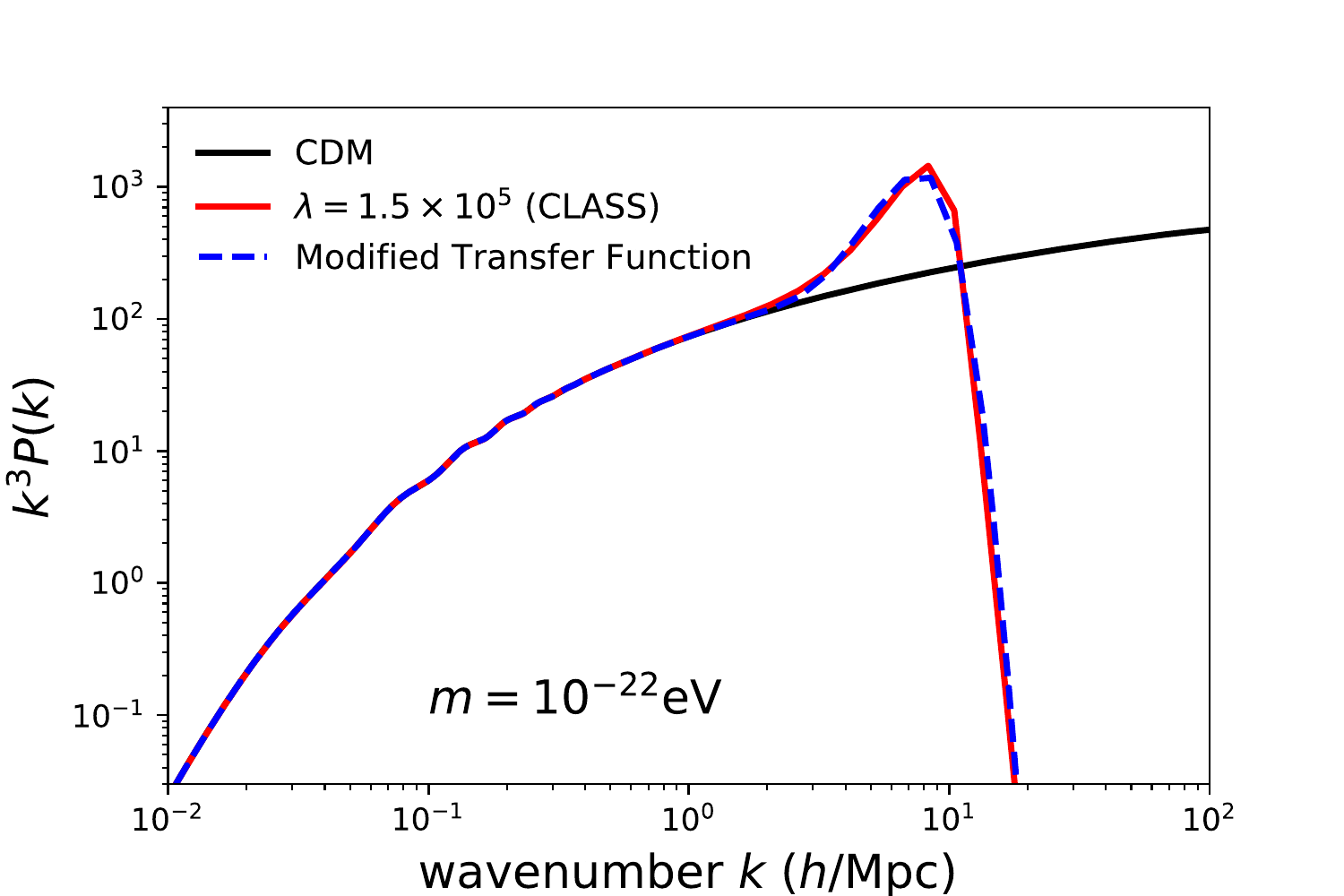}
        \label{fig:first_sub}   
    \caption{Dimensionless MPS for CDM (black), SFDM from \textsc{class} (blue), SFDM from the modified transfer function (red). The parameters of the axion field are $m_{\phi}=10^{-22}$eV and $\lambda=1.5\times 10^5$. The parameters of the modified transfer function~\eqref{mod_tf} are given by $\left\lbrace \alpha\, , \beta\, , \gamma\, , \delta\, , \eta\, , \nu \right\rbrace = \left\lbrace 0.087\, , 5.4\, , -2.7\, , 5\times 10^{-4}\, , 6.6\, , 0.2\right\rbrace$.}
    \label{mps_tf}
\end{figure}
\end{center}

Additionally, Figure~\ref{mps_tf} shows the dimensionless MPS for CDM (black line), SFDM with tachyonic instability obtained from \textsc{class} (red line), and the MPS obtained from the transfer function~\eqref{mod_tf} (blue line) via the expression $P_{SFDM}(k) = T^2(k) P_{CDM}(k)$. It can be seen that it is possible to reproduce the bump at the cut--off scale with the modified transfer function~\eqref{mod_tf}. Therefore, since the Lyman-$\alpha$ likelihood implemented in the latest version of \textsc{Monte Python} is only
applicable for models that can be described by the $\left\lbrace \alpha\, , \beta\, , \gamma \right\rbrace$ parameterization given in Eq.~\eqref{tf}, it is then clear that such likelihood is not appropriate to constrain our model. 

\bibliographystyle{unsrt}
\bibliography{bib}

\end{document}